%% file: main.tex
\documentclass[%
 reprint,
 superscriptaddress,
 amsmath,amssymb,
 aps,
 prl,
]{revtex4-2}
\usepackage{graphicx}% Include figure files
\usepackage{dcolumn}% Align table columns on decimal point
\usepackage{bm}% bold math
\usepackage{siunitx}
\usepackage{lineno}
\usepackage{xcolor}
\usepackage{comment}
\usepackage{adjustbox}
\usepackage[utf8x]{inputenc}
\usepackage{ucs}
\usepackage{lineno}
\usepackage{multirow}
\usepackage[normalem]{ulem}
\usepackage{subfigure}

\begin{document}
%\linenumbers
\title{Search for light dark matter with ionization signals in the PandaX-4T Experiment}

\def\shKeyLab{School of Physics and Astronomy, Shanghai Jiao Tong University, Key Laboratory for Particle Astrophysics and Cosmology (MoE), Shanghai Key Laboratory for Particle Physics and Cosmology, Shanghai 200240, China}
\def\BUAA{School of Physics, Beihang University, Beijing 102206, China}
\def\BUAALab{Beijing Key Laboratory of Advanced Nuclear Materials and Physics, Beihang University, Beijing, 102206, China}
\def\zzu{School of Physics and Microelectronics, Zhengzhou University, Zhengzhou, Henan 450001, China}
\def\USTClab{State Key Laboratory of Particle Detection and Electronics, University of Science and Technology of China, Hefei 230026, China}
\def\USTCdep{Department of Modern Physics, University of Science and Technology of China, Hefei 230026, China}
\def\BUAALab{International Research Center for Nuclei and Particles in the Cosmos \& Beijing Key Laboratory of Advanced Nuclear Materials and Physics, Beihang University, Beijing 100191, China}
\def\pku{School of Physics, Peking University, Beijing 100871, China}
\def\YaLongSD{Yalong River Hydropower Development Company, Ltd., 288 Shuanglin Road, Chengdu 610051, China}
\def\IAP{Shanghai Institute of Applied Physics, Chinese Academy of Sciences, 201800 Shanghai, China}
\def\CHEPpku{Center for High Energy Physics, Peking University, Beijing 100871, China}
\def\SDUdep{Research Center for Particle Science and Technology, Institute of Frontier and Interdisciplinary Science, Shandong University, Qingdao 266237, Shandong, China}
\def\SDUlab{Key Laboratory of Particle Physics and Particle Irradiation of Ministry of Education, Shandong University, Qingdao 266237, Shandong, China}
\def\UMD{Department of Physics, University of Maryland, College Park, Maryland 20742, USA}
\def\TDLee{Tsung-Dao Lee Institute, Shanghai Jiao Tong University, Shanghai, 200240, China}
\def\MESJTU{School of Mechanical Engineering, Shanghai Jiao Tong University, Shanghai 200240, China}
\def\SYU{School of Physics, Sun Yat-Sen University, Guangzhou 510275, China}
\def\SYUSFI{Sino-French Institute of Nuclear Engineering and Technology, Sun Yat-Sen University, Zhuhai, 519082, China}
\def\SYUzhuhai{School of Physics and Astronomy, Sun Yat-Sen University, Zhuhai, 519082, China}
\def\NKU{School of Physics, Nankai University, Tianjin 300071, China}
\def\FDU{Key Laboratory of Nuclear Physics and Ion-beam Application (MOE), Institute of Modern Physics, Fudan University, Shanghai 200433, China}
\def\USST{School of Medical Instrument and Food Engineering, University of Shanghai for Science and Technology, Shanghai 200093, China}
\def\SJTUSC{Shanghai Jiao Tong University Sichuan Research Institute, Chengdu 610213, China}
\def\SPEIT{SJTU Paris Elite Institute of Technology, Shanghai Jiao Tong University, Shanghai, 200240, China}
\affiliation{\shKeyLab}
\author{Shuaijie Li}\affiliation{\TDLee}
\author{Mengmeng Wu}\affiliation{\SYU}
\author{Abdusalam Abdukerim}\affiliation{\shKeyLab}
\author{Zihao Bo}\affiliation{\shKeyLab}
\author{Wei Chen}\affiliation{\shKeyLab}
\author{Xun Chen}\affiliation{\shKeyLab}\affiliation{\SJTUSC}
\author{Yunhua Chen}\affiliation{\YaLongSD}
\author{Chen Cheng}\affiliation{\SYU}
\author{Zhaokan Cheng}\affiliation{\SYUSFI}
\author{Xiangyi Cui}\affiliation{\TDLee}
\author{Yingjie Fan}\affiliation{\NKU}
\author{Deqing Fang}\affiliation{\FDU}
\author{Changbo Fu}\affiliation{\FDU}
\author{Mengting Fu}\affiliation{\pku}
\author{Lisheng Geng}\affiliation{\BUAA}\affiliation{\BUAALab}\affiliation{\zzu}
\author{Karl Giboni}\affiliation{\shKeyLab}
\author{Linhui Gu}\affiliation{\shKeyLab}
\author{Xuyuan Guo}\affiliation{\YaLongSD}
\author{Chencheng Han}\affiliation{\shKeyLab} %
\author{Ke Han}\affiliation{\shKeyLab}
\author{Changda He}\affiliation{\shKeyLab}
\author{Jinrong He}\affiliation{\YaLongSD}
\author{Di Huang}\affiliation{\shKeyLab}
\author{Yanlin Huang}\affiliation{\USST}
\author{Zhou Huang}\affiliation{\shKeyLab}
\author{Ruquan Hou}\affiliation{\SJTUSC}
\author{Xiangdong Ji}\affiliation{\UMD}
\author{Yonglin Ju}\affiliation{\MESJTU}
\author{Chenxiang Li}\affiliation{\shKeyLab}
\author{Jiafu Li}\affiliation{\SYU}
\author{Mingchuan Li}\affiliation{\YaLongSD}
\author{Shu Li}\affiliation{\MESJTU}
\author{Qing Lin}\email[Corresponding author: ]{qinglin@ustc.edu.cn}\affiliation{\USTClab}\affiliation{\USTCdep}
\author{Jianglai Liu}\email[Spokesperson: ]{jianglai.liu@sjtu.edu.cn}\affiliation{\shKeyLab}\affiliation{\TDLee}\affiliation{\SJTUSC}
\author{Xiaoying Lu}\affiliation{\SDUdep}\affiliation{\SDUlab}
\author{Lingyin Luo}\affiliation{\pku}
\author{Yunyang Luo}\affiliation{\USTCdep}
\author{Wenbo Ma}\affiliation{\shKeyLab}
\author{Yugang Ma}\affiliation{\FDU}
\author{Yajun Mao}\affiliation{\pku}

\author{Yue Meng}\email[Corresponding author: ]{mengyue@sjtu.edu.cn}\affiliation{\shKeyLab}\affiliation{\SJTUSC}
\author{Xuyang Ning}\affiliation{\shKeyLab}
\author{Ningchun Qi}\affiliation{\YaLongSD}
\author{Zhicheng Qian}\affiliation{\shKeyLab}
\author{Xiangxiang Ren}\affiliation{\SDUdep}\affiliation{\SDUlab}
\author{Nasir Shaheed}\affiliation{\SDUdep}\affiliation{\SDUlab}
\author{Changsong Shang}\affiliation{\YaLongSD}
\author{Xiaofeng Shang}\affiliation{\shKeyLab}
\author{Guofang Shen}\affiliation{\BUAA}
\author{Lin Si}\affiliation{\shKeyLab}
\author{Wenliang Sun}\affiliation{\YaLongSD}
\author{Andi Tan}\affiliation{\UMD}
\author{Yi Tao}\affiliation{\shKeyLab}\affiliation{\SJTUSC}
\author{Anqing Wang}\affiliation{\SDUdep}\affiliation{\SDUlab}
\author{Meng Wang}\affiliation{\SDUdep}\affiliation{\SDUlab}
\author{Qiuhong Wang}\affiliation{\FDU}
\author{Shaobo Wang}\affiliation{\shKeyLab}\affiliation{\SPEIT}
\author{Siguang Wang}\affiliation{\pku}
\author{Wei Wang}\affiliation{\SYUSFI}\affiliation{\SYU}
\author{Xiuli Wang}\affiliation{\MESJTU}
\author{Zhou Wang}\affiliation{\shKeyLab}\affiliation{\SJTUSC}\affiliation{\TDLee}
\author{Yuehuan Wei}\affiliation{\SYUSFI}
\author{Weihao Wu}\affiliation{\shKeyLab}
\author{Jingkai Xia}\affiliation{\shKeyLab}
\author{Mengjiao Xiao}\affiliation{\UMD}
\author{Xiang Xiao}\affiliation{\SYU}
\author{Pengwei Xie}\affiliation{\TDLee}
\author{Binbin Yan}\affiliation{\shKeyLab}
\author{Xiyu Yan}\affiliation{\SYUzhuhai}
\author{Jijun Yang}\affiliation{\shKeyLab}
\author{Yong Yang}\affiliation{\shKeyLab}
\author{Yukun Yao}\affiliation{\shKeyLab}
\author{Zhengyun You}\affiliation{\SYU}
\author{Chunxu Yu}\affiliation{\NKU}
\author{Jumin Yuan}\affiliation{\SDUdep}\affiliation{\SDUlab}
\author{Ying Yuan}\affiliation{\shKeyLab}
\author{Zhe Yuan}\affiliation{\FDU} %
\author{Xinning Zeng}\affiliation{\shKeyLab}
\author{Dan Zhang}\affiliation{\UMD}
\author{Minzhen Zhang}\affiliation{\shKeyLab}
\author{Peng Zhang}\affiliation{\YaLongSD}
\author{Shibo Zhang}\affiliation{\shKeyLab}
\author{Shu Zhang}\affiliation{\SYU}
\author{Tao Zhang}\affiliation{\shKeyLab}
\author{Yang Zhang}\affiliation{\SDUdep}\affiliation{\SDUlab}
\author{Yingxin Zhang}\affiliation{\SDUdep}\affiliation{\SDUlab} %
\author{Yuanyuan Zhang}\affiliation{\TDLee}
\author{Li Zhao}\affiliation{\shKeyLab}
\author{Qibin Zheng}\affiliation{\USST}
\author{Jifang Zhou}\affiliation{\YaLongSD}
\author{Ning Zhou}\email[Corresponding author: ]{nzhou@sjtu.edu.cn}\affiliation{\shKeyLab}\affiliation{\SJTUSC}
\author{Xiaopeng Zhou}\affiliation{\BUAA}
\author{Yong Zhou}\affiliation{\YaLongSD}
\author{Yubo Zhou}\affiliation{\shKeyLab}

\collaboration{PandaX Collaboration}
\noaffiliation
\date{\today}% It is always \today, today,

\begin{abstract}
We report the search results of light dark matter through its interactions with shell electrons and nuclei, using the commissioning data from the PandaX-4T liquid xenon detector. 
Low energy events are selected to have an ionization-only signal between 60 to 200 photoelectrons, corresponding to a mean nuclear recoil %energy from \sout{1.19} 0.77 to \sout{2.85} 2.54\,keV and
energy from 0.77 to 2.54\,keV and electronic recoil energy from 0.07 to 0.23\,keV. 
%\sout{0.10} 0.07 to \sout{0.28} 0.23\,keV. 
With an effective exposure of 0.55\,tonne$\cdot$year, we set the most stringent limits within a mass range from 40\,$\rm{MeV/c^2}$ to 10\,$\rm{GeV/c^2}$ for point-like dark matter-electron interaction, 100\,MeV/c$^2$ to 10\,GeV/c$^2$ for dark matter-electron interaction via a light mediator, and 3.2 to 4\,$\rm{GeV/c^2}$ for dark matter-nucleon spin-independent interaction. For DM interaction with electrons, our limits are 
%significant parameter space for sub-GeV dark matter 
%produced either by freeze-out or freeze-in mechanisms in the early Universe 
%has been probed the first time, 
closing in on the parameter space predicted by the freeze-in and freeze-out mechanisms in the early Universe.

\end{abstract}

\maketitle

% quick functions
\newcommand{\mwba}[1]{\textcolor{violet}{#1}}
\newcommand{\mwbd}[1]{\textcolor{violet}{\sout{#1}}}

%The existence of dark matter is established by overwhelming gravitational evidence in cosmological and astronomical observations~\cite{BERTONE2005279review}, but its physics nature remains a mystery. The weakly-interacting-massive-particle (WIMP) is one of the well-motivated dark matter candidates, as proposed from many theories beyond the Standard Model of particle physics.
Dark matter (DM) direct detection experiments are being carried out worldwide to detect possible interactions between the DM and baryonic matter~\cite{Liu:2017drfPandaXII,Undagoitia_2015_DMreview}. 
DM particles within a mass range from about 5~$\rm{GeV/c^2}$ to 10~$\rm{TeV/c^2}$ have been extensively searched for via the recoil of atomic nucleus~\cite{Tan:2016dizPandaXII, Aprile:2018dbl_xenon1tNR, Akerib:2016vxi_LUXNR, Agnes:2018fwg_darksideNR,Ajaj:2019imk,Agnese:2013jaa,Jiang:2018pic,Abdelhameed:2019hmk,Amole:2017dex,meng2021dark}. 
Such DM may have been naturally frozen-out in the early Universe and become the thermal relic~\cite{Kolb:1990vq}. 
%and the explored parameter space expands significantly in recent years~\cite{Tan:2016dizPandaXII, Aprile:2018dbl_xenon1tNR, Akerib:2016vxi_LUXNR, Agnes:2018fwg_darksideNR,Ajaj:2019imk,Agnese:2013jaa,Jiang:2018pic,Abdelhameed:2019hmk,Amole:2017dex,meng2021dark}. 
Lighter DM particles are also well-motivated theoretically. In addition to the thermal freeze-out, they could also be produced slowly in non-equilibrium along the evolution of Universe (freeze-in)~\cite{Hall:2009bx}.
% {\color{red}(Yang: should explain the feature difference between freeze-in and freeze-out mechanisms using one or two sentences, to make it more readable.)} 
Detecting them with conventional techniques, however, becomes more difficult as the recoil energy is much suppressed. 
Many low threshold techniques have been developed in recent years, enabling experimental searches for the light DM scatterings with nuclei and with shell electrons~\cite{pandaxiiewimp,PhysRevD.85.076007DMmodel,PhysRevLett.109.021301, PhysRevLett.123.181802,  Emken_2019,Essig:2015cda,Essig:2017kqs}.
%However, the energy of recoil signals from light WIMP particles is too low to overcome the detection threshold of conventional techniques, the corresponding sensitivity is much weaker. 
In this letter, we report a dedicated low threshold search for the light DM particles with ionization-only signals using the commissioning data of PandaX-4T.

The PandaX-4T experiment~\cite{meng2021dark,hongguang,Qian_2022,He_2021,Chen_2021,Zhao_2021,XiuliIndium2021} is located in the B2 experimental hall of the China Jinping Underground Laboratory (CJPL). The central detector is a dual-phase Time Projection Chamber (TPC) containing an active 
cylindrical sensitive target with 3.7\,$\rm{tonnes}$ of liquid xenon (LXe). An energy deposition in LXe produces prompt scintillation photons ($S1$) and ionized electrons in the liquid. Ionized electrons are drifted under the electrical field defined by the cathode and gate grid located at the bottom and top of the LXe, respectively. They are extracted and amplified by a stronger field in between the gate and anode across the liquid level, producing electroluminescence photons ($S2$) proportional to the number of ionized electrons. 
Both $S1$ and $S2$ signals are detected by two arrays of Hamamatsu R11410-23 photomultiplier tubes (PMTs) located at the top and bottom of the TPC~\cite{meng2021dark}. 
For each scintillation photon and ionized electron being produced, the average detection efficiencies are measured to be 9\% and 90\%, respectively. Therefore, conventional requirement of $S1$-$S2$ pairs inevitably leads to significant efficiency loss for low energy events. Conversely, ionization-only events open up a low energy window down to just a few ionized electrons.

Similar to Ref.~\cite{b8_paper}, a blind analysis is performed on the sets 2 and 4-5 of 
the commissioning data of PandaX-4T~\cite{b8_paper}. 
Data set 3 was removed due to the high micro-discharge noises (MD) from the electrodes.
Events with unpaired $S2$ (US2), i.e. no accompanying $S1$ greater than 2 photoelectrons (PE), are defined as the candidates.
The requirement on the accompanying $S1$ has an efficiency of $>$90\% for all the DM models probed in this analysis.
Data of approximately eight days are randomly selected to validate the signal selection and background composition.
The signal selection consists of three steps, the signal reconstruction, the data quality cuts, and the region-of-interest (ROI) selection. 
 The signal reconstruction refers to the identification and reconstruction of the $S2$-like signals from the raw data. 
To correctly reconstruct the $S2$ signal, PMT hits belonging to an $S2$-like signal are clustered by taking into account the diffusion of the electrons during their drift.  
Different from previous analyses~\cite{meng2021dark, b8_paper}, the $S2$ clustering algorithm in this analysis is modified to be solely based on the $S2$ charge and width due to the lack of vertical position reconstruction. 
The data quality cuts are developed based on the calibration data to remove noises and unphysical events, including pile-ups of single-electron events, the background originated from electrodes, events happening in the gaseous region, and MD events.
To collect enough statistics of small $S2$ signals, the secondary $S2$s from the double scattering (DS) events of the $^{241}$Am-Be and DD calibration are selected. 
The secondary $S2$ is required to precede the main $S2$ of the DS events, in order to ensure the purity of the DS sample.
%based on the calibration data
%The data quality cuts are developed based on the distribution of secondary $S2$ from double scattering (DS) events of $^{241}$Am-Be and DD calibration to collect higher statistics of smaller $S$2 signals in region of interested (ROI). 
%The $S2$ from a DS event is required to precede the main $S2$, in order to ensure the purity of the DS sample. 
Similar to Ref.~\cite{b8_paper}, the data quality cuts are defined based on the $S2$ horizontal position reconstruction quality, the top-bottom charge ratio, the signal waveform shape, %$S2$ pulse shape, isolation time, ($\mathcal{F}_{S2}$) 
the veto PMT charge, and the afterglow veto. 
 %\sout{(``baseline selection'')}
% \textcolor{red}{In addition, we develop one additional cut based on the charge of the most fired PMT (maximum charge PMT? should be a ratio?? need some details and a word such as "To remove ..., ").} \textcolor{blue}{to remove the MDs which happen on one single PMT.}
%The single electrons pile-up events, electrodes background, background happened in the gaseous region and micro-discharge (MD) events could be removed with above cuts. 
The data quality cuts (cut0) are further optimized by maximizing signal-to-background ratio, where the background is assumed to consist of solely the so-called cathode background (see later text).
For the signal waveform shape cuts, in particular, a selection on S2 width with S2-dependent upper and lower boundaries is adopted, which suppresses the cathode background most significantly.
% \textcolor{red}{Especially for the signal waveform shape cuts, a strict selection on S2 width with an $S2$-dependent upper and lower boundary was used. This cut suppressed the cathode background most significantly.
% }

%\sout{(``optimized selection'')} 
% Not clear what this means, and very red herring. Later on it appears "optimized cut" is good in removing MD events, leading one to suspect whether optimized cut was defined post-blinding ...% before unblinding the data ROI.
%The selections before and after optimization are called baseline and optimized selections, respectively.
The $S2$ ROI of this analysis is set to be between 60 to 200\,PE.
The lower boundary of the ROI is determined to avoid the apparent high background rate at very low $S2$, and the upper boundary is set so that the loss of the sensitivity to DM search is negligible (no more than 5\% for all the DM models investigated). 
% Problematic and meaning not clear. For ER/NR DM models in this paper, the ROI effiencies are all different ...
The ``afterglow'' veto cut~\cite{b8_paper} is applied to remove the exposure time with high afterglow $S2$ rate. The total effective live time of the selected US2 data is 64.7 days. 
The same radial cut as in Ref.~\cite{meng2021dark} is also applied, leading to a fiducial mass of $3.1 \pm 0.1$ tonnes, thus, a total effective exposure of 0.55\,tonne$\cdot$year.

\begin{figure}[htbp]
    \includegraphics[width=0.45\textwidth]{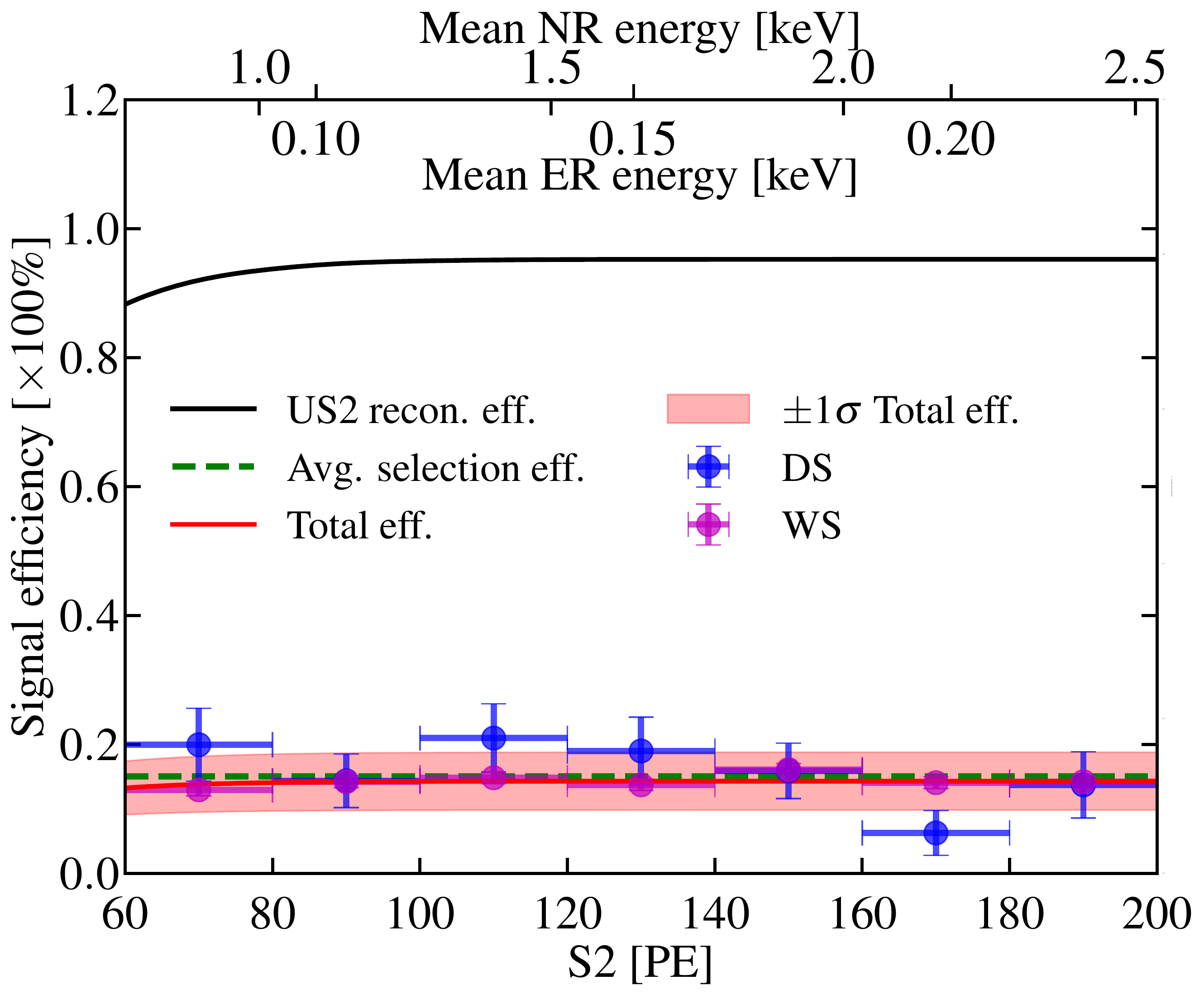}
    \caption{\label{fig:eff} 
    Total efficiency broken down to US2 reconstruction efficiency (black solid line), and data quality selection efficiencies (green) evaluated using the DS (blue) and WS (magenta) samples. 
    With the data quality selection efficiency modeled as a constant (green dashed line), the total efficiency is shown with  $\pm$1$\sigma$ band in light red.
    The mean NR and ER energy scales in the ROI are indicated on the top axis, which are evaluated based on flat energy spectra.
    %  Selection efficiency broken down to US2 reconstruction efficiency (black curve) and data quality selection efficiencies evaluated using the DS (blue) and WS (magenta) data samples.
    % The mean selection efficiency is modelled as a constant (green dashed lines) with 
    % $\pm$1$\sigma$ band in light green. The corresponding mean NR and ER energy scales are indicated on the top axis. 
    %\textcolor{red}{(QL: Remove the baseline selection and correspoinding labels of baseline and optimized selections. updated)}
    }
\end{figure}

For events located in the ROI, the signal efficiency, separated into the US2 reconstruction efficiency and data quality selection efficiency, is shown in Fig.~\ref{fig:eff}.
%The trigger efficiency is 100\% since the PandaX-4T data acquisition utilizes a trigger-less readout scheme~\cite{yang2022readout}. %which is fully efficient %trigger efficiency of 100\% for the US2 events in the ROI.
The US2 reconstruction efficiency is evaluated through dedicated waveform simulation (WS), as described in Ref.~\cite{b8_paper}. 
It starts to drop below about 80\,PE with an efficiency of about 90\% at 60\,PE, with negligible systematic uncertainty.
%Its systematic uncertainty due to the WS's uncertainty is negligible.
%The US2 reconstruction efficiency is evaluated through dedicated waveform simulation (WS) \textcolor{blue}{and accidental coincidence probability Ref.~\cite{b8_paper}.}
%\textcolor{red}{The systematic uncertainty of the reconstruction efficiency is about xx\% (@LSJ)}.
The data quality selection efficiency is estimated %as the fraction of survived events with all selections applied 
using the DS and the WS samples~\cite{b8_paper}, and no significant $S2$ dependence is observed. 
Therefore, the average of two methods is taken as the nominal value, and the standard deviation is taken as the systematic uncertainty ($\sim$31\%).
%(@LSJ need re-estimate)}.
%and we take the constant average as the final efficiency for the data quality selections.
%We take the standard deviations of the efficiencies obtained using the DS and WS samples as the systematic uncertainty, which is about 23\%.
%The PandaX-4T data acquisition utilizes a trigger-less readout scheme~\cite{yang2022readout}, which is fully efficient %trigger efficiency of 100\% 
%for the US2 events in the ROI. Fig.~\ref{fig:eff} shows the selection efficiencies with baseline and optimized selections, and US2 reconstruction efficiency.

Prior to the unblinding of the data, three background compositions are evaluated.
%Before unblinding the data, we consider three background components: 
The electronic recoil (ER) background is primarily due to beta decays of the internal radioactivities such as tritium and $^{222}$Rn.   %\textcolor{red}{(why no top/bottom PMT etc???)} 
The nuclear recoil (NR) background is produced by the solar $^8$B neutrino elastic scattering off xenon nuclei (CE$\nu$NS) and the neutron background. 
The nominal rates and energy spectra of these backgrounds are the same as those in Ref.~\cite{b8_paper}.
Due to the lack of $S1$, the US2 data are also contaminated by the background 
emerging from the radioactivities in the cathode or on its surface, exhibiting a signature of $S2$ with large width due to diffusion effects \footnote{Such background can also emerge from the gate electrode, but is significantly suppressed by our data selection based on the $S2$ width}.
%are from previous best-fit of data in the dark matter search~\cite{meng2021dark} and from Ref.~\cite{ruppin2014complementarity}, respectively. 
%\textcolor{red}{
%\sout{The cathode background is measured from the paired $S1-S2$ events originating from the cathode at the bottom of the TPC sensitive volume, which is then multiplied by the ratio of US2 to paired $S1-S2$ events.}
% The cathode background is estimated by comparing the distributions of the US2 data and the selected cathode events with a paired $S1$, which provide a controlled cathode background sample.
%} 
%$S2$ signals with large width %\sout{($w_{cum}$, the width of waveform enclosing $10\%-90\%$ cumulative charge)} 
%are dominated by the cathode background, which provides a controlled cathode sample for the estimation of the cathode background in the US2 data. 
% Details of this estimate is provided in the supplementary material. 
To obtain features of the cathode background, tagged cathode events in $S1$-$S2$ pairs with characteristic vertical positions are selected. 
% The same baseline quality cuts of US2 are implemented for the $S2$ signals. In addition, the $S1$ charge is required to be smaller than 100~PE to mimic the US2 cathode background. 
For $S1$ less than 100\,PE, the $S2$ distribution of the selected cathode events is found to be independent of $S1$, and is therefore taken as the shape of the US2 cathode background. 
The rate of the US2 cathode background is obtained by scaling 
the tagged cathode events with $S2$ in the ROI.
%the rate of  the selected cathode events in the S2 ROI.
% The $S2$ charge and width distributions of the selected controlled cathode events and the US2 events are shown in Fig.~\ref{fig:cathode}.
The scaling constant is the ratio between the US2 events and tagged cathode events in a side-band with $S2$ from 200 to 350\,PE and $S2$-width from 2.5 to 4.5\,$\mu$s.
The systematic uncertainty is estimated to be 25\%, by varying the side-band region.
% In total, 12 control regions are chosen with various $S2$ charge, width, and $S1$ charge ranges, to evaluate the normalization of the US2 cathode background and its systematic uncertainty. 

%and their predictions before unblinding and statistical interpretation are shown in Table~\ref{tab:bkgcomp}.
%The cathode background is estimated based on the control regions. The details are in the supplementary. We use the cathode events outside of ROI as primary controlled samples for estimating the cathode background. The uncertainty is estimated based on the difference of cathode background with different controlled samples in ($S1$, $S2$, $S2$ width) bins. The $S2$ spectrum of the cathode background can be found in Fig.~\ref{fig:bkg}.

\begin{table}[htp]
    \centering
    \begin{tabular}{c|c|c}
    \hline\hline
    & Nominal & Best-fit  \\ %&\textcolor{red}{Observed}\\
    \hline\hline
    Cathode     & %45.0&$\pm$&8.0
    41.6$\pm$10.6  &  63.9$\pm$9.1 \\
    MD          & 6.9$^{+9.0}$             
    &  17.7$\pm$5.3 \\
    Solar $\nu$ & 10.8$\pm$3.7  &  11.7$\pm$3.6 \\
     ER          & 2.3$\pm$0.6   &  2.5$\pm$0.5 \\
    Neutron     & 0.1$\pm$0.1   &  0.1$\pm$0.1 \\
    \hline
    Total       &  61.7$_{-11.2}^{+14.4}$  & 95.8$\pm$11.3 \\
    % &105 \\
    \hline\hline
    \end{tabular}
    \caption{
    Nominals and background-only best-fits of the background components in the US2 candidates.
    %in $S2$ range from 60 to 200\,PE in data with 64.7-day effective live time of this analysis, 
    %\textcolor{red}{(QL: Need to update with the new fitting procedure: two-step unblinding.)}
    }
    \label{tab:bkgcomp} 
\end{table}%

%test flat Cathode
% \begin{table}[htp]
%     \centering
%     \begin{tabular}{c|c|c}
%     \hline\hline
%     & Nominal & Best-fit  \\ %&\textcolor{red}{Observed}\\
%     \hline\hline
%     Cathode     & %45.0&$\pm$&8.0
%     41.4$\pm$10.5  &  66.0$\pm$9.7 \\
%     MD          & 6.9$^{+9.0}$             
%     &  16.8$\pm$5.2 \\
%     Solar $\nu$ & 10.8$\pm$3.7  &  11.4$\pm$3.5 \\
%      ER          & 2.3$\pm$0.6   &  2.4$\pm$0.5 \\
%     Neutron     & 0.1$\pm$0.1   &  0.1$\pm$0.1 \\
%     \hline
%     Total       &  61.7$_{-11.2}^{+14.4}$  & 96.8$\pm$11.8 \\
%     % &105 \\
%     \hline\hline
%     \end{tabular}
%     \caption{
%     Nominals and background-only best-fits of the background components in the US2 candidates.
%     %in $S2$ range from 60 to 200\,PE in data with 64.7-day effective live time of this analysis, 
%     %\textcolor{red}{(QL: Need to update with the new fitting procedure: two-step unblinding.)}
%     }
%     \label{tab:bkgcomp} 
% \end{table}%

\begin{figure}[!t]
    \includegraphics[width=0.5\textwidth]{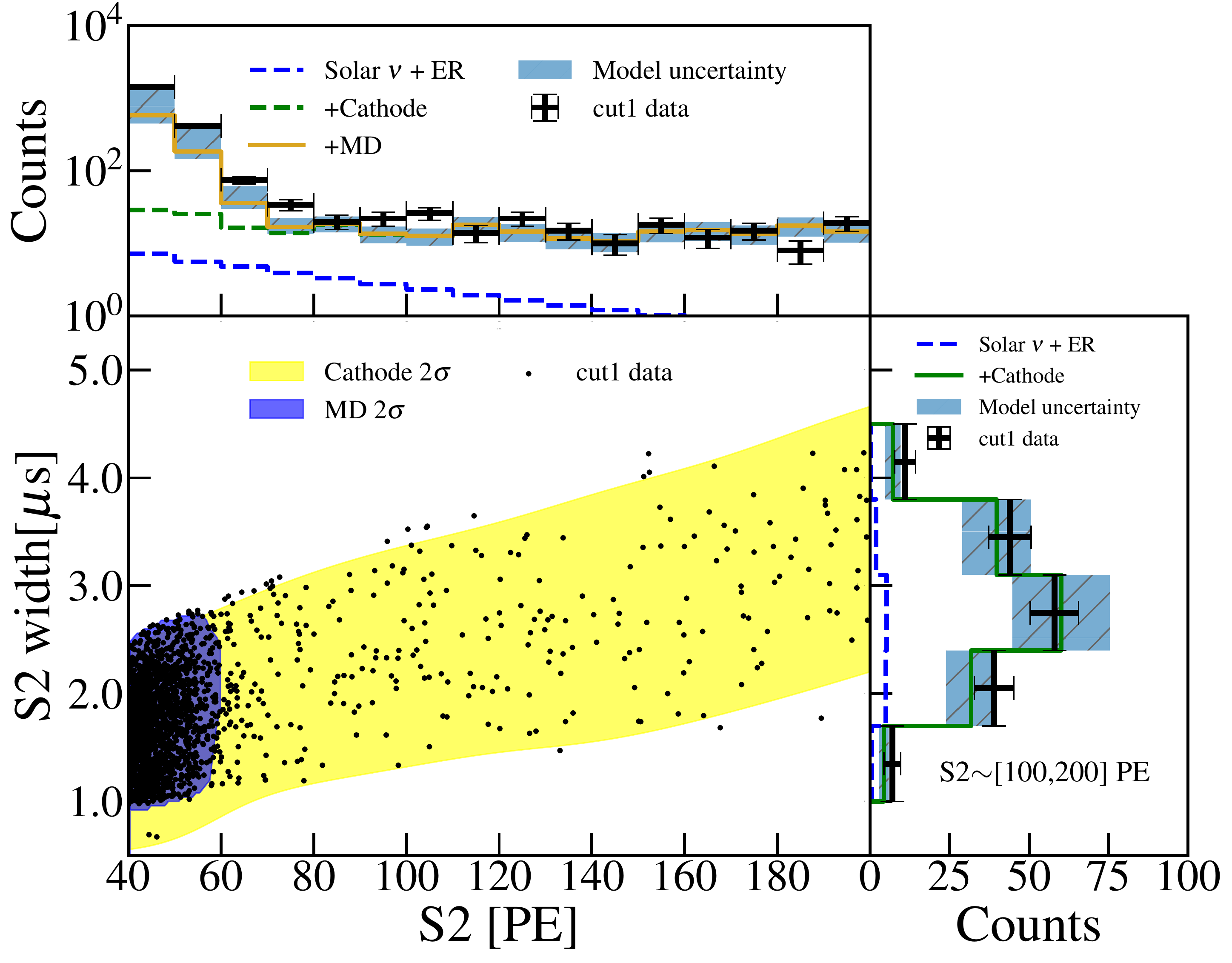}

    \caption{
    Main: $S2$ width vs. $S2$ for events under cut1, together with 
    expected contours of the cathode background (yellow=2$\sigma$) and MD background (violet=2$\sigma$)). Top and right panels:
    projections to $S2$ (all events) and to $S2$ width for $S2$ within 100 to 200\,PE, respectively. Various stacked background components are indicated by the legend, and the background uncertainty is represented by the shaded regions. 
    % \textcolor{violet}{(@LSJ WMM try add 40-60PE, keep only the 1$\sigma$ contours of the MD and cathode background. Remove the signal 1$\sigma$ contour. Change "cut1-cut0" back to "cut1".)}
    %The blue dashed, green dashed, and golden solid lines represent the stacked background distributions with the solar $\nu$ + ER background, the cathode background, and the MD background added in sequence, respectively.
    %The shaded regions of the top and right panels give the $\pm$1$\sigma$ confidence region of the rates of the background model.
    }
    \label{fig:fit_control_region}
\end{figure}

A two-step unblinding procedure is carried out on our data to ensure good control over background.
In the first step, we define a set of loosened data quality cuts on the US2 width, shape, top/bottom charge ratio etc., and unblind the complementary events (cut1 data) between the loosened cut and cut0, which is more sensitive to instrumental background. The event distribution in cut1 is shown in Fig.~\ref{fig:fit_control_region}.
Shown on the right panel is the distribution of $S2$-width, which is particularly sensitive to cathode background, for events between 100 and 200\,PE (nearly no MD contribution). 
The agreement between the background prediction and the data is good.
%Data of the COMP set have higher background rate and its $S2$ charge and width are used to further test the background model.
At the very low-$S2$ region below 80\,PE, a clear excess is observed with a charge and width distributions consistent with the MD 
in set 3 (excluded already). The rate of the excess also 
varies with data-taking periods, 1.7 times higher in data sets 4-5 than that in set 2, indicating a residual level of MD after set 3. 
Therefore, a MD background component is added to 
the background model for sets 4-5 only, with the rate estimated from the difference between sets 4-5 and set 2 in an $S2$ side-band region from 40 to 60\,PE \footnote{
Under cut1, $S2$s in set 2 are stretched due to a slightly different $S2$ gain of this dataset.}, and the shape taken from set 3. 
%$S2$ charge distribution taken from the MD background in set~3, 
%The rate of the MD background is conservatively estimated by fitting the difference of the event rates before and after the set~3 in the COMP set with the $S2$ in range of [60, 80]\,PE.
The comparison between the cut1 data and background prediction is shown in the upper panel of Fig.~\ref{fig:fit_control_region}.
Our nominal MD background in S2 [60, 80] PE undershoots the observed cut1 rate by 130\%, which may be consistent with a residual MD background component in set 2, or a constant DM signal degenerate with MD background. We assign +130\% as an asymmetrical systematic uncertainty for the MD rate in sets 4-5.

%for signals in [100, 200]\,PE, where the lower cut is to eliminate MD contribution.  
%In addition, the cathode background is further tested using the $S2$ width distribution with $S2$ in [100, 200]\,PE.
% The rate of MD background, along with the rate of the cathode background, is further constrained using the $S2$ charge and width distributions in the control region. 
% The US2 events from set~3 
% %are dominated by the MD background, 
% provide a reliable template of the MD background.
% To minimize the possible DM signal interference in the MD background modelling, the complementary data set between the ones with the baseline and optimized selections is taken as the controlled MD sample, and the difference of event rate between data before and after set~3 is fitted with the MD template. 
%It is assumed that the increasing of event rate between data before and after Set 3 is caused by the MD. We take the complementary set of data between baseline and optimized selections as the control samples for estimating the MD background rate. Data from Set 3 are taken as the template of distribution for the MD background. 
The MD prediction under the final cut0 is then set by the MD expectation in Fig.~\ref{fig:fit_control_region} scaled by the ratio of events between cut0 and cut1 in data set 3.
%The scalings from the COMP set to the US2 data for the MD backgrounds is determined as the ratio of the aforementioned set~3 data after the corresponding data selections.
The obtained rate and the associated uncertainty of the MD background are listed in Table~\ref{tab:bkgcomp} (column ``nominal'').
%The systematic uncertainty of the MD background model consists of two components: the difference of the MD model to the COMP data in the distribution of $S2$ and the selection (different $S2$ ranges) of the COMP set. 
%The overall uncertainty is estimated to be 29\%.
%31\%. 
% The predicted MD background in $S2$ from 60 to 200\,PE with optimized selection is $24.2\pm7.5$. 
% More information on the MD background modelling is available in the supplementary material. 
%The scalings from the COMP set to the US2 data for the MD backgrounds is determined as the ratio of the aforementioned set~3 data after the corresponding data selections.
%The obtained rate and the associated uncertainty of the MD background are listed in Table~\ref{tab:bkgcomp} (column ``prior'').
%The $S2$ charge distributions of the MD and cathode background with the final data selection are also shown in Fig.~\ref{fig:bkg}. 

\begin{figure}[!th]
    \includegraphics[width=0.45\textwidth]{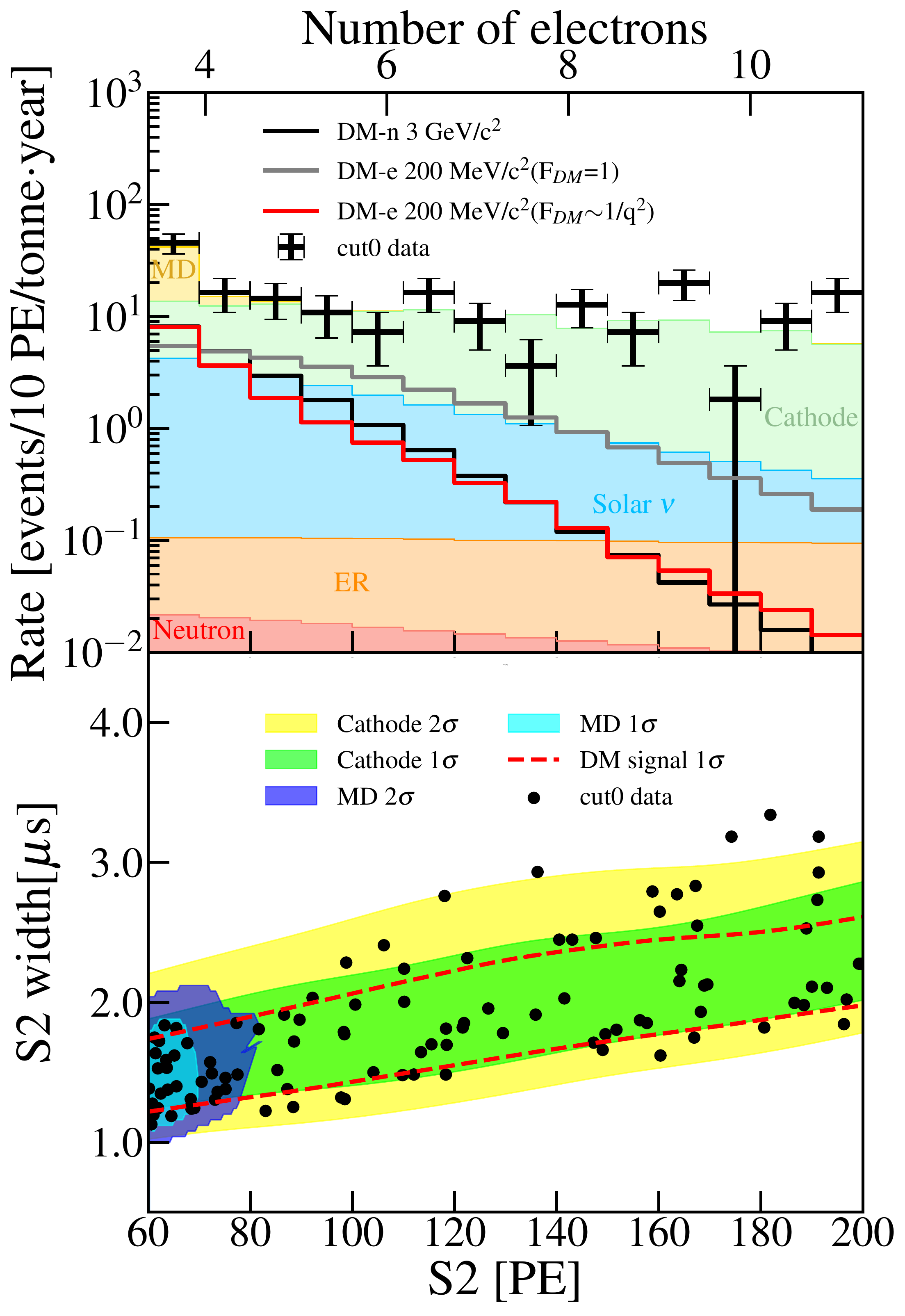}
    \caption{
    \label{fig:bkg}
    Top: candidates (cut0) and stacked background components %baseline (top) and optimized (bottom)
    from the background-only best fit. %\textcolor{red}{A $\chi^2$ based on the data and the statistical uncertainties in comparison to the summed background is 20.7 (NDF=13).} 
    %data selections, rates of which come from the best-fits of the statistical interpretation under the background-only hypothesis.
    %The red, orange, blue, green, yellow regions show the background contributions from the neutron, ER, solar $\nu$, cathode, and MD, respectively.
    %The black markers represent the data $S2$ spectra.
    %The black solid and dashed lines show the expected $S2$ spectra from DM-nucleon spin-independent scatterings with mass of 3 and 5\,GeV/c$^2$, respectively, assuming a cross section of 10$^{-43}$cm$^2$. 
    Expected distributions for DM interactions are also overlaid and indicated by the legend, with assumed cross section of $10^{-43}$cm$^2$ (DM-nucleon), 10$^{-41}$cm$^2$ (DM-e with F$_{\textrm{DM}}$=1), 10$^{-36}$cm$^2$ (DM-e with F$_{\textrm{DM}}  \sim 1 / q^2$).
    % \textcolor{violet}{(@LSJ WMM Change style to be same as Fig.2, with two panels. Top panel gives the S2 spectra as the old one, bottom panel 
    Bottom: final candidate in $S2$ width vs. $S2$ together with the
    expected contours of the DM signal and the cathode and MD background (see legend).}
    
\end{figure}%
% \begin{figure}[htp]
%     \centering
%     \quad
%     \subfigure[]{
%     \includegraphics[width=0.5\textwidth]{e-wimp/md_width.pdf}
%     }
%     \quad
%     \subfigure[]{
%     \includegraphics[width=0.5\textwidth]{e-wimp/cathode_width.pdf}
%     }
%     \caption{
%     \textcolor{red}{
%     The $S2$ width distributions in the control region with $S2$ in [60,80]\,PE (top panel) and [100, 200]\,PE (bottom panel), respectively, dominated by the MD and cathode background.
%     The black solid lines give the best-fit models.
%     }
%     }
%     \label{fig:fit_control_region}
% \end{figure}

% \begin{figure*}[!btp]
%     \centering
%     \includegraphics[width=0.9\textwidth]{e-wimp/S2only_ER_upperlimits_2.pdf}
%     \caption{
%     The 90\% C.L. upper limits on cross sections of the point-like interaction with form-factor F$_{\textrm{DM}}$=1 (left panel) and light mediator with $\si{F_{DM}}\sim 1/q^2$ (right panel) with the constant-$W$ model~\cite{essig2017new} and the NEST 2.3.6 model~\cite{PhysRevD.103.012002} are shown. Red solid line represent the upper limits of this work. The constraints given by other experiments~\cite{aprile2019light, agnes2022er,essig2017new,pandaxiiewimp,PhysRevLett.123.181802,Abramoff_2019} are also overlaid. The cross-sections considering DM freeze-in and freeze-out abundance are plotted as well~\cite{Essig:2015cda}.
%     }
%     \label{fig:limit}
% \end{figure*}

\begin{figure*}[!btp]
    \centering
    \includegraphics[width=0.9\textwidth]{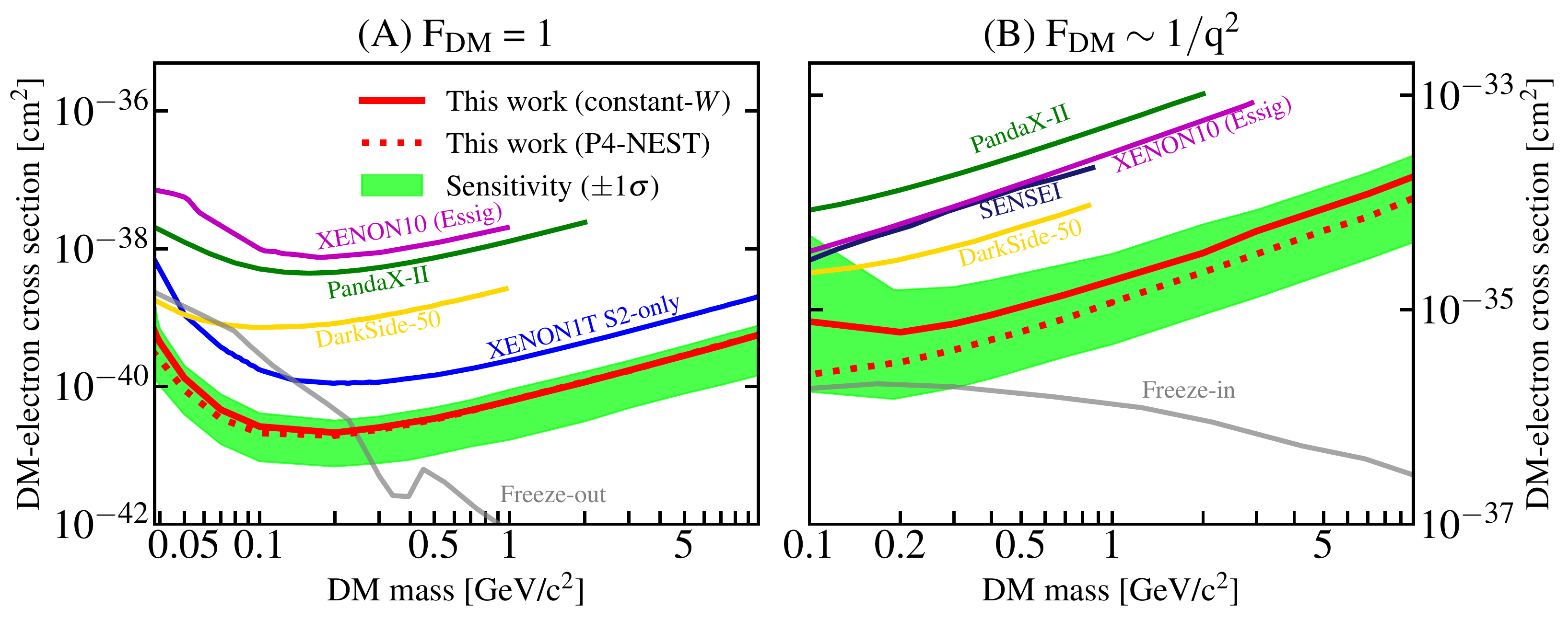}
    \caption{
    The 90\% C.L. upper limits on DM-electron cross sections of the point-like interaction with $\si{F_{DM}}=1$ (left panel) and with the light mediator with $\si{F_{DM}}\sim 1/q^2$ (right panel)
    using the constant-$W$ model (red solid, official results) and P4-NEST model (red dashed), as well as the green $\pm$1$\sigma$ sensitivity band. For comparison, results from other experiments~\cite{aprile2019light, agnes2022er,essig2017new,pandaxiiewimp,PhysRevLett.123.181802,Abramoff_2019}, as well as theoretical predictions from DM vector-portal freeze-in and freeze-out mechanisms ($\Omega h^2$=0.12)~\cite{Essig:2015cda}, 
    are also overlaid. 
    }
    \label{fig:limit}
\end{figure*}

\begin{figure}[!h]
    \centering
    \includegraphics[width=0.45\textwidth]{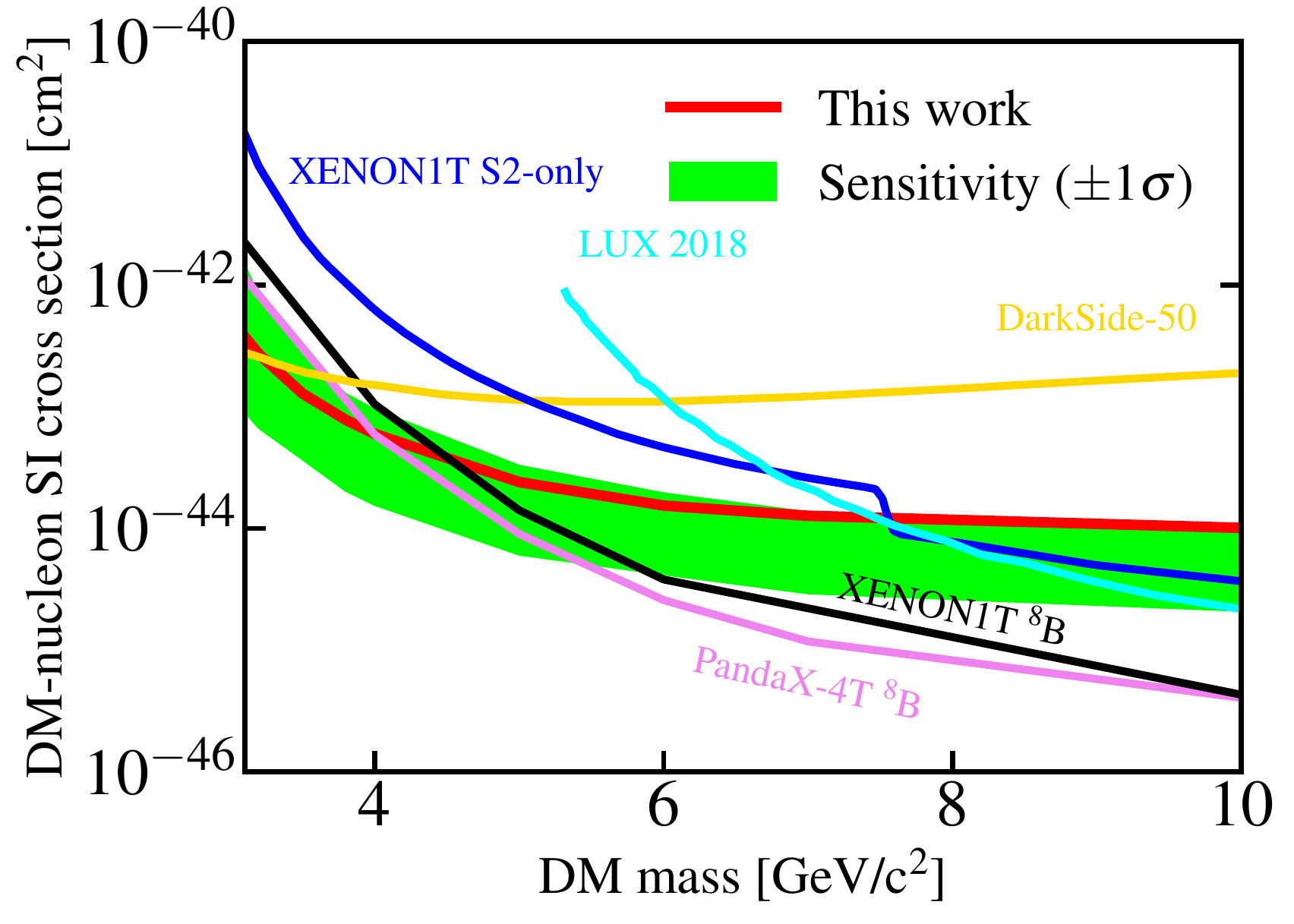}
    \caption{
    The 90\% C.L. exclusion limit on the spin-independent DM-nucleon cross section with $\pm$1$\sigma$ sensitivity, together with 
    results from other work~\cite{b8_paper, aprile2021search, akerib2017results,agnes2022_nr,essig2017new}.}
    \label{fig:limit_nr}
\end{figure}%

% \textcolor{red}{Appears to be out of place. I tried to move it right before the fit, but realize that this paragraph also covers the ER/NR background modeling, without which one could not discuss the MD excess. But the ER/NR DM signal should come later, I think.}

%We adopt the same signal response models as in Refs.~\cite{meng2021dark, b8_paper} 
%, essig2017new} 
%to obtain the corresponding $S2$ distribution of the ERs and NRs for the background and DM %signals.
%for background and DM signals. 
As in Refs.~\cite{meng2021dark, b8_paper}, our NR signal model %\textcolor{red}{NR} 
follow the construction of the NEST package~\cite{szydagis2018noble}, with the parameters obtained from a fit to the PandaX-4T calibration data~\cite{meng2021dark}, and 
extrapolate to our ROI (P4-NEST). 
%Due to the lack of measurements of very low-energy recoils in PandaX-4T, we extrapolate the signal model parameters in our ROI (P4-NEST). 
The systematic uncertainty is dominated by the charge yield uncertainty in NEST~\cite{szydagis2022noble} for a recoil energy below $\sim$5\,keV NR energy. 
%($\sim$0.4\,keV ER energy).
%In the WIMP-nucleon analysis, the signal response model takes NEST parameterization~\cite{szydagis2018noble}, and fits to the ER and NR calibration data in higher energy range~\cite{meng2021dark} and extrapolates to ROI. Due to the lack of measurements of very low-energy recoils, the NEST parameterization has systematic uncertainties~\cite{szydagis2022noble} on charge yields in xenon, which is taken into account in this analysis. It is the dominant systematic uncertainty as the energy gets below $\sim$5\,keV NR energy (equivalent to approximately 0.4\,keV ER energy). 
The ER energy scale in the lowest region of the ROI also has large uncertainty as the lowest 
ever calibration point in LXe is 0.186\,keV~\cite{LUX:2017ojt}.
Therefore, for the ER events, a more conservative constant-$W$ model~\cite{essig2017new} is chosen as the nominal model.
The ROI corresponds to the mean energy range from 0.77 to 2.54\,keV for NRs (P4-NEST), and 0.07 to 0.23\,keV for ERs (constant-$W$). The solar $\nu$, ER and neutron background can then be estimated under these models (Table~\ref{tab:bkgcomp}). The surface background is negligible within the FV cut.
%to conservatively estimate the number of primary ionized electrons, as well as to be consistent with other analysis. 
%In general, the constant-$W$ model predicts smaller charge yield in comparison with the P4-NEST model. 
%The $S2$ spectra 
%%obtained using constant-$W$ model 
%for the DM-nucleus and DM-electron scatterings with typical masses are overlaid in Fig.~\ref{fig:bkg}. %Their calculations follow the same as in Ref.~\cite{meng2021dark} and~\cite{essig2017new}.\\ 
% More details of the two charge yield models can be found in the supplementary material.

\begin{table}[htp]
    \centering
    \begin{tabular}{ll|c|l}
    \hline\hline
    Nuisance parameters & & Stdev. & Estimated by \\
    \hline\hline
    Data sel. eff. & $\delta_\epsilon$ & 0.31 & DS vs. WS\\
    Sig. model rate & $\delta_{\textrm{s}}f_i$ & $f_i$ & NEST uncert.\\
    Cathode bkg. rate & $\delta_{\textrm{cat}}$ & 0.25 & ROI side-band  \\
    MD bkg. rate & $\delta_{\textrm{MD}}$  & $^{+1.3}_{-0.0}$  & cut1 data \\
    \hline\hline
    \end{tabular}
    \caption{
    Summary of the standard deviations of the nuisance parameters (nominal values all at 0) used in the statistical interpretation (see text). $f_i$ is an correlated energy-dependent fractional uncertainty of the DM signal rate in each $S2$ bin.}
    \label{tab:sys}
\end{table}%

The second step unblinding happens after all background nominals in the ROI are set. In total 105 events are observed in cut0, shown in Fig.~\ref{fig:bkg}.
%, where the best-fit background contributions from the MD, cathode, solar $\nu$, ER and neutron events are also overlaid. 
%The $S2$ spectra for the DM-nucleus and DM-electron scatterings with typical masses are also shown.
% Statistical interpretation of the US2 data is performed based on a profile likelihood ratio (PLR) method~\cite{baxter2021recommended} \textcolor{red}{using a two-sided test statistics}.
Statistical interpretation of the US2 data is performed based on a two-sided profile likelihood ratio (PLR) method~\cite{baxter2021recommended}.
% Besides the US2 data in the ROI after the optimized selection, we also add the US2 data after the baseline selection with $S2$ from 100 to 200\,PE as the controlled US2 data to constrain the cathode background. %\textcolor{red}{(!!QL: it seems we have too may controlled samples in the paper.)}
% \textcolor{red}{Not sure what it means. The definition of "optimized cut" mentioned nothing about cathode background.}
The binned likelihood of this analysis is defined as:
\begin{equation}
    \centering
    \mathcal{L} =  G(\delta_\epsilon)G(\delta_s) G(\delta_{\textrm{cat}})G(\delta_{\textrm{MD}}) \prod \limits_i \frac{\lambda_i^{N_i}}{N_i !} e^{-\lambda_i} ,
    % \prod \limits_j \frac{\lambda_j^{N_j}}{N_j !} e^{-\lambda_j},
    \label{eq:likelihood}
\end{equation}
\noindent
where $N_i$ and $\lambda_i$ are the observed and predicted events, respectively, in the $i$-th bin in $S2$. 
$\delta_\epsilon$, $\delta_s$, $\delta_{\textrm{cat}}$, and $\delta_{\textrm{MD}}$ are the nuisance parameters 
%with nominal values of 0 
%which have mean values of 0 and 
corresponding to the systematic uncertainties of the data selection efficiency, the DM signal rate, the cathode background rate, and the MD background rate, respectively, constrained by Gaussian terms $G$ (see also Table~\ref{tab:sys}).
%All the 1$\sigma$ values of the nuisance parameters and corresponding systematic uncertainties are described in Table.~\ref{tab:sys}.
%The parameter $\delta_s$ is factored together with the fractional uncertainty of the signal rate $f_i$ which depends on the $S2$ and DM signal spectrum, similar to Ref.~\cite{b8_paper}, and has {\color{red}{typical uncertainties are from 1\% up to 37\% depending on $S$2 range}}
%\sout{of 34.8\% and 11.9\%, respectively, at $S2$=100\,PE for the DM signals with the masses of 40 and} 
%with the DM mass 200\,MeV/c$^2$ assuming a DM-electron scatter with F$_{\textrm{DM}}$=1.
The expected events $\lambda_i$ can be written as:
\begin{equation}
    \centering
    \begin{aligned}
\lambda_i  = & N_i^{\chi} (1+\delta_{\textrm{s}} f_i) (1+\delta_\epsilon) +  N_i^{\textrm{cathode}} (1+\delta_{\textrm{cat}}) \\
& + N_i^{\textrm{MD}} (1+\delta_{\textrm{MD}})+ N_i^{\textrm{others}} , \\
 \end{aligned}
    \label{eq:lambda}
\end{equation}%
\noindent where $N_i^{\chi}$, $N_i^{\textrm{cathode}}$, $N_i^{\textrm{MD}}$, and $N_i^{\textrm{others}}$ are the nominal events for the DM signals, cathode background, MD background, and other background (solar $\nu$, ER, and neutrons), respectively.
% \textcolor{red}{
% The DM signal and background models for set~2 and sets~4-5 are separately generated due to their slightly different detector conditions, and then summed up as the final models.
% }
The DM signal and background models for set~2 and sets~4-5 are generated separately according to their detector conditions and then summed up.
The parameter $\delta_s$ is factored together with a fractional shape uncertainty $f_i$ which depends on actual DM signal spectrum, similar to Ref.~\cite{b8_paper}. For example, for a point-like 200\,MeV/c$^2$ DM-electron interaction, $f_i$ varies from 1\% to 37\% from 60 to 200 PE. 
%{\color{red}{typical uncertainties are from 1\% up to 37\% depending on $S$2 range}}
%\sout{of 34.8\% and 11.9\%, respectively, at $S2$=100\,PE for the DM signals with the masses of 40 and} 
%with the DM mass 200\,MeV/c$^2$ assuming a DM-electron scatter with F$_{\textrm{DM}}$=1.
% $f_i$ is the fractional systematic uncertainty of the DM signal rate which depends on the $S2$ and DM signal spectrum, similar to Ref.~\cite{b8_paper}, and, for example, has typical values of xx and xx, respectively, for the DM masses of 20 and 200\,MeV/c$^2$ assuming a DM-electron scatter with F$_{\textrm{DM}}$=1.
The background-only best-fit rates of the background components are summarized in Table.~\ref{tab:bkgcomp}. 
An upward shift is observed in the cathode background, nevertheless within two standard deviations from the nominal, which is conservative in the case of setting exclusion limits.
No significant excess is observed above expected background, therefore our data are cast into DM exclusion limits. The 2$\sigma$ upward shift in the fitted cathode background implies an underestimation of the background, but is nevertheless conservative in the case of limit setting. Likewise, our nominal MD background and its asymmetric uncertainty also lead to a more conservative limit.
%Therefore, our data are interpreted as exclusion limits on theinteraction cross sections. 

Three benchmark models are considered in this analysis: the DM-electron elastic scatterings with a heavy mediator (the DM form factor $\si{F_{DM}}=1$) and a light mediator (%\sout{$\si{F_{DM}}=\alpha^2 m_e^2/q^2$} 
$\si{F_{DM}}\sim 1/q^2$, where $q$ is the momentum transfer)~\cite{essig2017new}, and the DM-nucleon spin-independent (SI) scattering. 
%the DM-electron elastic scattering and DM-nucleon spin-independent cross-sections%~\cite{essig2017new}, 
% \textcolor{red}{The exclusion limits on the scattering cross-sections at 90\% C.L., which are shown in Fig.~\ref{fig:limit} and Fig.~\ref{fig:limit_nr}, are about 1.3$\times$10$^{-43}$cm$^2$ and 2.1$\times$10$^{-41}$cm$^2$, respectively, with DM masses of 3.5\,GeV/c$^2$ for DM-nucleon SI scattering and of 200\,MeV/c$^2$ for DM-electron scattering with a heavy mediator.  }
The exclusion limits on the scattering cross-sections at 90\% C.L. are shown in Fig.~\ref{fig:limit} and Fig.~\ref{fig:limit_nr}. To set the scale, the limits are about 1.3$\times$10$^{-43}$~cm$^2$ (3.5\,GeV/c$^2$) for the DM-nucleon SI scattering and 2.1$\times$10$^{-41}$~cm$^2$ (200\,MeV/c$^2$) for DM-electron scattering with a heavy mediator. 
The obtained results have provided the most stringent constraints for the DM-electron interactions with mass in range of 40\,MeV/c$^2$ to 10\,GeV/c$^2$ with F$_{\textrm{DM}}$=1, and 100\,MeV/c$^2$ to 10\,GeV/c$^2$ with 
%\sout{F$_{\textrm{DM}} = \alpha^2 m_e^2 / q^2$} 
F$_{\textrm{DM}}\sim 1/q^2$, and for DM-nucleon SI interactions in the DM mass range of 3.2 to 4\,GeV/c$^2$. 
% \textcolor{red}{The sensitivities for DM models with masses lower the reported ranges show significantly large values due to the large systematic uncertainty on the charge yields.
% Thus, the results with lower DM masses are not reported.
% }
Results with lower DM masses are not reported, as the sensitivity band grows significantly  due to the large systematic uncertainty on the charge yields.
% The results have ruled out the most likely parameter space with DM mass between approximately 0.04 and 0.25\,GeV/c$^2$, for the freeze-out mechanism (DM-electron scattering with F$_{\textrm{DM}}$=1).
% If P4-NEST signal response model is used, the results can also exclude the most probable parameter space with the DM mass between about 0.1 and 0.4\,GeV/c$^2$ for the freeze-in mechanism (DM-electron scattering with F$_{\textrm{DM}}\sim 1/q^2$).
Our exclusions on DM-electron interactions represent a significant step-forward in the field. Under the assumption of vector portal interactions (e.g. the dark photon as the mediator), our results challenge the freeze-out mechanism for DM mass range from 0.04 to 0.25 GeV/c$^2$ with F$_{\textrm{DM}}$=1, and are closing in on the freeze-in prediction with F$_{\textrm{DM}}\sim 1/q^2$, assuming such light DM provides the entire DM abundance.
%that light DMs freeze-out and -in via the vector portal (e.g. the dark photon as the mediator)~\cite{Essig:2015cda}, our results show that the freeze-out mechanism is not preferred for DM mass range from 0.04 to 0.25 GeV/c$^2$ with F$_{\textrm{DM}}\sim 1/q^2$, assuming such light DM provides the all DM abundance.
%If the P4-NEST signal response model is used, the PandaX-4T result didn't support the freeze-in mechanism for DM mass ranging from 0.1 to 0.3\,GeV/c$^2$ with F$_{\textrm{DM}}\sim 1/q^2$, assuming such light DM provides the all DM abundance. 
% (For freeze in, increasing this interaction leads to more dark matter but in freeze-out, increasing this coupling leads to less dark matter.)

In summary, a blind analysis using ionization-only data from the PandaX-4T commissioning run 
is carried out to search for light DM interactions with xenon nuclei and atomic electrons.
We have lowered the $S2$ threshold to 60\,PE, equivalent to a mean NR energy of about 0.77\,keV and ER energy of about 0.07\,keV. All background components in the ROI are understood and well constrained.
%Dedicated background models are developed to address the new background components emerged with the ionization-only data and the lowered threshold, compared to Ref.~\cite{meng2021dark}. 
With an effective exposure of 0.55\,tonne$\cdot$year, no significant excess is observed above background. 
Thus, we have obtained the leading constraints on the DM-electron cross-sections with the DM mass in the range of 40\,MeV/c$^2$ to 10\,GeV/c$^2$ for a heavy mediator,
%with form-factor F$_{\textrm{DM}}$=1
and 100\,MeV/c$^2$ to 10\,GeV/c$^2$ for a light mediator, respectively, and on the DM-nucleon SI cross-sections within the DM mass range from 3.2 to 4\,GeV/c$^2$.
%with form-factor F$_{\textrm{DM}} = \alpha^2 m_e^2 / q^2$, respectively.
% PandaX-4T continues taking more physics data, and \sout{is expected} \textcolor{red}{hopes} to improve the sensitivity by another order of magnitude with a 6-tonne-year exposure.
PandaX-4T is taking more physics data and working to suppress the background further, aiming to further improve the sensitivity with a 6-tonne-year total exposure.
\\
\input{acknowledgement.tex}

\bibliographystyle{apsrev4-2.bst}
\bibliography{apssamp}

\begin{figure*}[htp]
    \centering
    \includegraphics[width=0.9\textwidth]{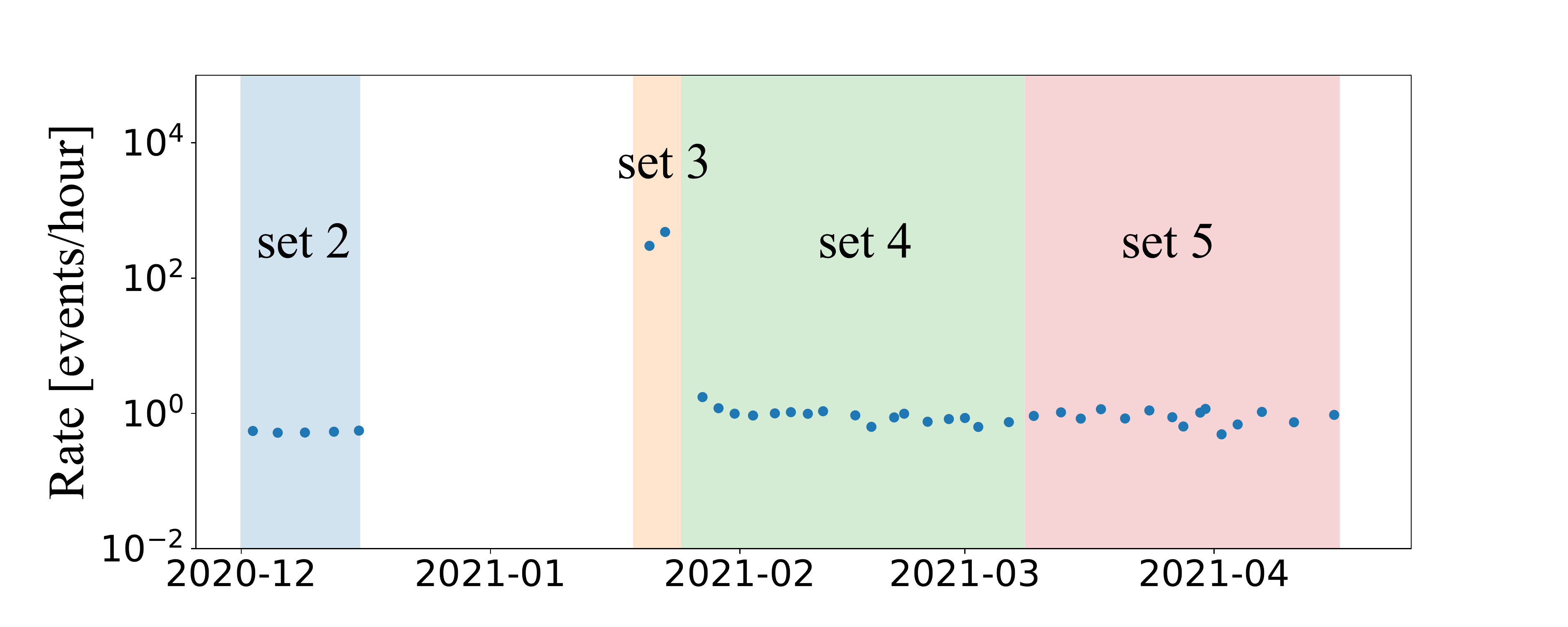}
    \caption{\label{fig:evolution} 
    The time evolution of the $S2$ (40-60 PE) event rate under cut1. The set 3 is dominated by the MD background. 
    The average rate of set~4 and set~5 are about 1.7 times higher than that of set~2.
    }
\end{figure*}%

\begin{figure*}[htp]
    \centering
    \includegraphics[width=0.75\textwidth]{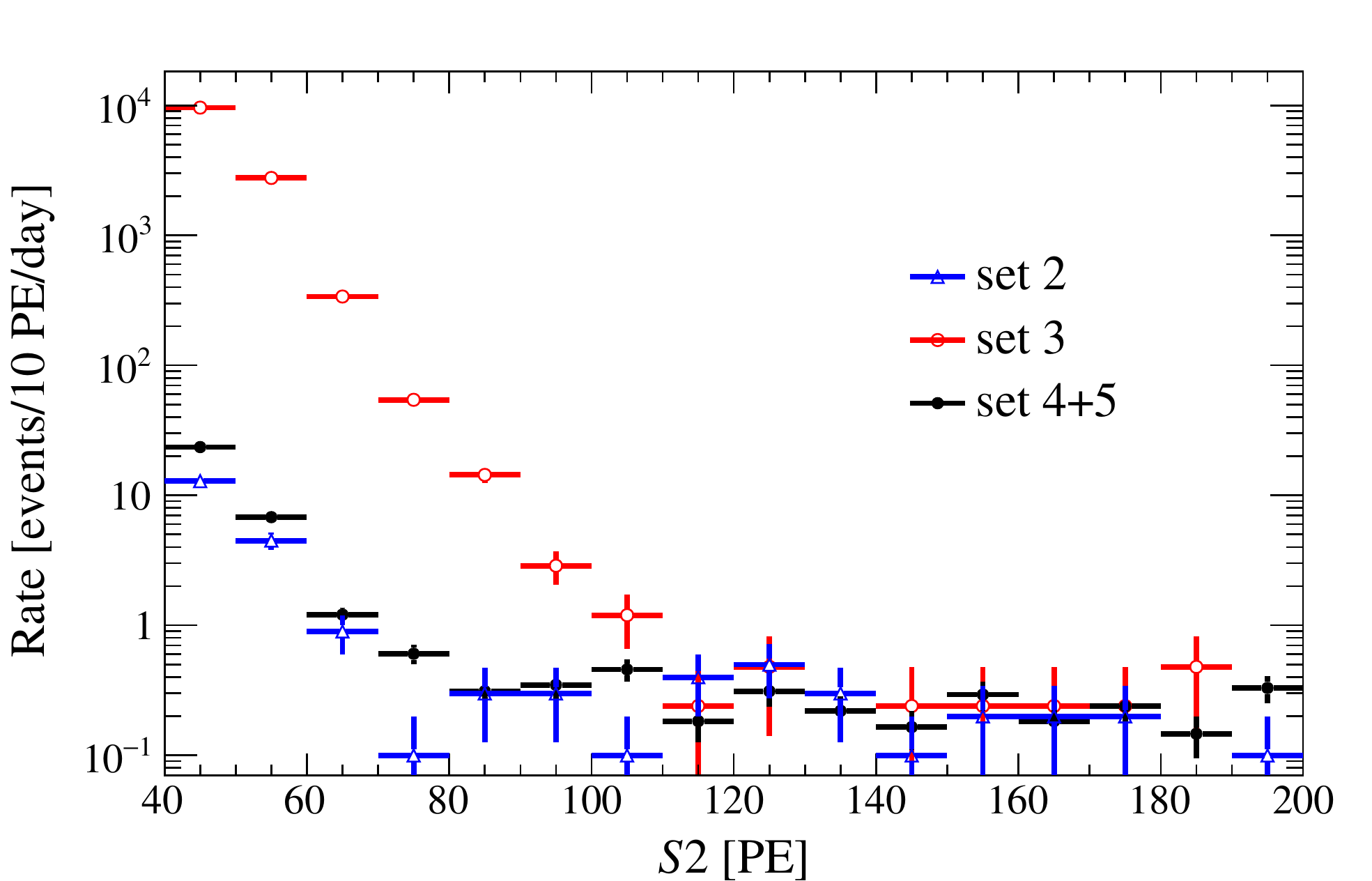}
    \caption{\label{fig:s2_rate} 
    The $S2$ spectra of set 2 (blue dots), set 3 (red circles), and set 4+5 (black dots).
    The $S2$s in the set~2 are scaled so that the $S2$ gain is consistent with set~4+5 when performing MD background estimation.
    % \textcolor{red}{@LSJ @WMM Remove set~2, and change "scaled set2 " to "set2".}
    % \textcolor{red}{(No need to label "unscaled")}
    }
\end{figure*}%

\begin{figure*}[!btp]
    \centering
    \includegraphics[width=0.75\textwidth]{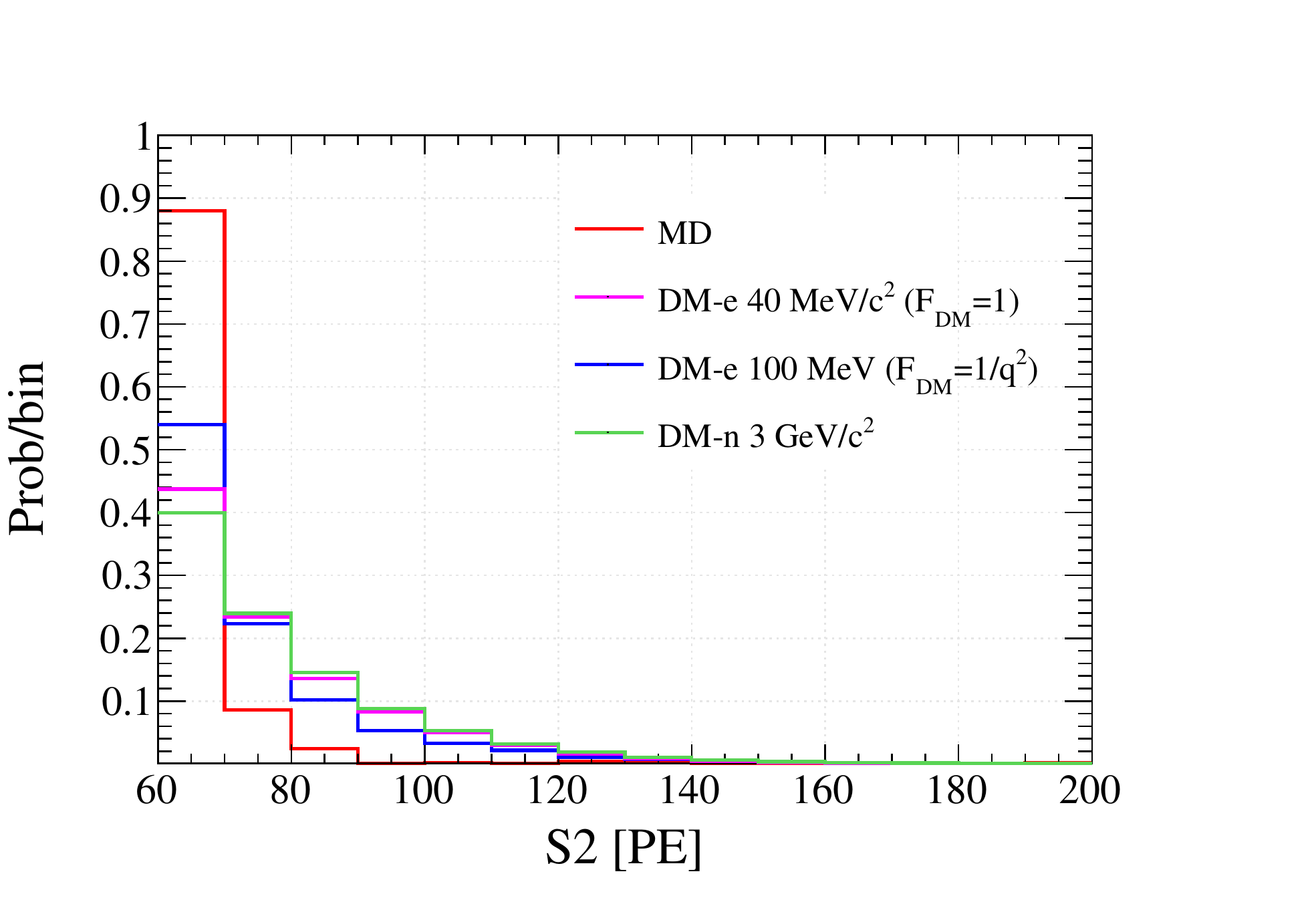}
    \caption{
    The comparison of the $S2$ spectra of MD (red), DM-e 40 MeV/c$^2$ (F$_{DM}=1$) (purple), DM-e 100 MeV/c$^2$ (F$_{DM}$=1/q$^2$) (blue), and DM-n 3 GeV/c$^2$ (green).
    }
    \label{fig:s2_rate} 
\end{figure*}%

\end{document}

%% file: acknowledgement.tex
% !TEX root = ../main.

% \section{Acknowledgement}

We would like to thank Matthew Szydagis for useful discussions concerning NEST model uncertainty.
This project is supported in part by grants from National Natural Science
Foundation of China (Nos. 12090061, 12005131, 11905128, 11925502, 12222505, 11835005), a grant from the Ministry of Science and Technology of China (No. 2016YFA0400301),
and by the Office of Science and
Technology, Shanghai Municipal Government (grant No. 18JC1410200). We thank supports from Double First Class Plan of
the Shanghai Jiao Tong University. 
We also thank the sponsorship from 
%the Chinese Academy of Sciences Center for Excellence in Particle
%Physics (CCEPP), 
the Hongwen Foundation in Hong Kong, Tencent
Foundation in China, and Yangyang Development Fund. Finally, we thank the CJPL administration and
the Yalong River Hydropower Development Company Ltd. for
indispensable logistical support and other help.

%% file: main.bbl
\begin{thebibliography}{41}%
\makeatletter
\providecommand \@ifxundefined [1]{%
 \@ifx{#1\undefined}
}%
\providecommand \@ifnum [1]{%
 \ifnum #1\expandafter \@firstoftwo
 \else \expandafter \@secondoftwo
 \fi
}%
\providecommand \@ifx [1]{%
 \ifx #1\expandafter \@firstoftwo
 \else \expandafter \@secondoftwo
 \fi
}%
\providecommand \natexlab [1]{#1}%
\providecommand \enquote  [1]{``#1''}%
\providecommand \bibnamefont  [1]{#1}%
\providecommand \bibfnamefont [1]{#1}%
\providecommand \citenamefont [1]{#1}%
\providecommand \href@noop [0]{\@secondoftwo}%
\providecommand \href [0]{\begingroup \@sanitize@url \@href}%
\providecommand \@href[1]{\@@startlink{#1}\@@href}%
\providecommand \@@href[1]{\endgroup#1\@@endlink}%
\providecommand \@sanitize@url [0]{\catcode `\\12\catcode `\$12\catcode
  `\&12\catcode `\#12\catcode `\^12\catcode `\_12\catcode `\%12\relax}%
\providecommand \@@startlink[1]{}%
\providecommand \@@endlink[0]{}%
\providecommand \url  [0]{\begingroup\@sanitize@url \@url }%
\providecommand \@url [1]{\endgroup\@href {#1}{\urlprefix }}%
\providecommand \urlprefix  [0]{URL }%
\providecommand \Eprint [0]{\href }%
\providecommand \doibase [0]{https://doi.org/}%
\providecommand \selectlanguage [0]{\@gobble}%
\providecommand \bibinfo  [0]{\@secondoftwo}%
\providecommand \bibfield  [0]{\@secondoftwo}%
\providecommand \translation [1]{[#1]}%
\providecommand \BibitemOpen [0]{}%
\providecommand \bibitemStop [0]{}%
\providecommand \bibitemNoStop [0]{.\EOS\space}%
\providecommand \EOS [0]{\spacefactor3000\relax}%
\providecommand \BibitemShut  [1]{\csname bibitem#1\endcsname}%
\let\auto@bib@innerbib\@empty
%</preamble>
\bibitem [{\citenamefont {Liu}\ \emph {et~al.}(2017)\citenamefont {Liu},
  \citenamefont {Chen},\ and\ \citenamefont {Ji}}]{Liu:2017drfPandaXII}%
  \BibitemOpen
  \bibfield  {author} {\bibinfo {author} {\bibfnamefont {J.}~\bibnamefont
  {Liu}}, \bibinfo {author} {\bibfnamefont {X.}~\bibnamefont {Chen}},\ and\
  \bibinfo {author} {\bibfnamefont {X.}~\bibnamefont {Ji}},\ }\href
  {https://doi.org/10.1038/nphys4039} {\bibfield  {journal} {\bibinfo
  {journal} {Nature Phys.}\ }\textbf {\bibinfo {volume} {13}},\ \bibinfo
  {pages} {212} (\bibinfo {year} {2017})}\BibitemShut {NoStop}%
\bibitem [{\citenamefont {Undagoitia}\ and\ \citenamefont
  {Rauch}(2015)}]{Undagoitia_2015_DMreview}%
  \BibitemOpen
  \bibfield  {author} {\bibinfo {author} {\bibfnamefont {T.~M.}\ \bibnamefont
  {Undagoitia}}\ and\ \bibinfo {author} {\bibfnamefont {L.}~\bibnamefont
  {Rauch}},\ }\href {https://doi.org/10.1088/0954-3899/43/1/013001} {\bibfield
  {journal} {\bibinfo  {journal} {Journal of Physics G: Nuclear and Particle
  Physics}\ }\textbf {\bibinfo {volume} {43}},\ \bibinfo {pages} {013001}
  (\bibinfo {year} {2015})}\BibitemShut {NoStop}%
\bibitem [{\citenamefont {Tan}\ \emph {et~al.}(2016)\citenamefont {Tan} \emph
  {et~al.}}]{Tan:2016dizPandaXII}%
  \BibitemOpen
  \bibfield  {author} {\bibinfo {author} {\bibfnamefont {A.}~\bibnamefont
  {Tan}} \emph {et~al.} (\bibinfo {collaboration} {PandaX}),\ }\href
  {https://doi.org/10.1103/PhysRevD.93.122009} {\bibfield  {journal} {\bibinfo
  {journal} {Phys. Rev. D}\ }\textbf {\bibinfo {volume} {93}},\ \bibinfo
  {pages} {122009} (\bibinfo {year} {2016})}\BibitemShut {NoStop}%
\bibitem [{\citenamefont {Aprile}\ \emph {et~al.}(2018)\citenamefont {Aprile}
  \emph {et~al.}}]{Aprile:2018dbl_xenon1tNR}%
  \BibitemOpen
  \bibfield  {author} {\bibinfo {author} {\bibfnamefont {E.}~\bibnamefont
  {Aprile}} \emph {et~al.} (\bibinfo {collaboration} {XENON}),\ }\href
  {https://doi.org/10.1103/PhysRevLett.121.111302} {\bibfield  {journal}
  {\bibinfo  {journal} {Phys. Rev. Lett.}\ }\textbf {\bibinfo {volume} {121}},\
  \bibinfo {pages} {111302} (\bibinfo {year} {2018})}\BibitemShut {NoStop}%
\bibitem [{\citenamefont {Akerib}\ \emph
  {et~al.}(2017{\natexlab{a}})\citenamefont {Akerib} \emph
  {et~al.}}]{Akerib:2016vxi_LUXNR}%
  \BibitemOpen
  \bibfield  {author} {\bibinfo {author} {\bibfnamefont {D.}~\bibnamefont
  {Akerib}} \emph {et~al.} (\bibinfo {collaboration} {LUX}),\ }\href
  {https://doi.org/10.1103/PhysRevLett.118.021303} {\bibfield  {journal}
  {\bibinfo  {journal} {Phys. Rev. Lett.}\ }\textbf {\bibinfo {volume} {118}},\
  \bibinfo {pages} {021303} (\bibinfo {year} {2017}{\natexlab{a}})}\BibitemShut
  {NoStop}%
\bibitem [{\citenamefont {Agnes}\ \emph {et~al.}(2018)\citenamefont {Agnes}
  \emph {et~al.}}]{Agnes:2018fwg_darksideNR}%
  \BibitemOpen
  \bibfield  {author} {\bibinfo {author} {\bibfnamefont {P.}~\bibnamefont
  {Agnes}} \emph {et~al.} (\bibinfo {collaboration} {DarkSide}),\ }\href
  {https://doi.org/10.1103/PhysRevD.98.102006} {\bibfield  {journal} {\bibinfo
  {journal} {Phys. Rev. D}\ }\textbf {\bibinfo {volume} {98}},\ \bibinfo
  {pages} {102006} (\bibinfo {year} {2018})}\BibitemShut {NoStop}%
\bibitem [{\citenamefont {Ajaj}\ \emph {et~al.}(2019)\citenamefont {Ajaj} \emph
  {et~al.}}]{Ajaj:2019imk}%
  \BibitemOpen
  \bibfield  {author} {\bibinfo {author} {\bibfnamefont {R.}~\bibnamefont
  {Ajaj}} \emph {et~al.} (\bibinfo {collaboration} {DEAP}),\ }\href
  {https://doi.org/10.1103/PhysRevD.100.022004} {\bibfield  {journal} {\bibinfo
   {journal} {Phys. Rev. D}\ }\textbf {\bibinfo {volume} {100}},\ \bibinfo
  {pages} {022004} (\bibinfo {year} {2019})}\BibitemShut {NoStop}%
\bibitem [{\citenamefont {Agnese}\ \emph {et~al.}(2014)\citenamefont {Agnese}
  \emph {et~al.}}]{Agnese:2013jaa}%
  \BibitemOpen
  \bibfield  {author} {\bibinfo {author} {\bibfnamefont {R.}~\bibnamefont
  {Agnese}} \emph {et~al.} (\bibinfo {collaboration} {SuperCDMS}),\ }\href
  {https://doi.org/10.1103/PhysRevLett.112.041302} {\bibfield  {journal}
  {\bibinfo  {journal} {Phys. Rev. Lett.}\ }\textbf {\bibinfo {volume} {112}},\
  \bibinfo {pages} {041302} (\bibinfo {year} {2014})}\BibitemShut {NoStop}%
\bibitem [{\citenamefont {Jiang}\ \emph {et~al.}(2018)\citenamefont {Jiang}
  \emph {et~al.}}]{Jiang:2018pic}%
  \BibitemOpen
  \bibfield  {author} {\bibinfo {author} {\bibfnamefont {H.}~\bibnamefont
  {Jiang}} \emph {et~al.} (\bibinfo {collaboration} {CDEX}),\ }\href
  {https://doi.org/10.1103/PhysRevLett.120.241301} {\bibfield  {journal}
  {\bibinfo  {journal} {Phys. Rev. Lett.}\ }\textbf {\bibinfo {volume} {120}},\
  \bibinfo {pages} {241301} (\bibinfo {year} {2018})}\BibitemShut {NoStop}%
\bibitem [{\citenamefont {Abdelhameed}\ \emph {et~al.}(2019)\citenamefont
  {Abdelhameed} \emph {et~al.}}]{Abdelhameed:2019hmk}%
  \BibitemOpen
  \bibfield  {author} {\bibinfo {author} {\bibfnamefont {A.}~\bibnamefont
  {Abdelhameed}} \emph {et~al.} (\bibinfo {collaboration} {CRESST}),\ }\href
  {https://doi.org/10.1103/PhysRevD.100.102002} {\bibfield  {journal} {\bibinfo
   {journal} {Phys. Rev. D}\ }\textbf {\bibinfo {volume} {100}},\ \bibinfo
  {pages} {102002} (\bibinfo {year} {2019})}\BibitemShut {NoStop}%
\bibitem [{\citenamefont {Amole}\ \emph {et~al.}(2017)\citenamefont {Amole}
  \emph {et~al.}}]{Amole:2017dex}%
  \BibitemOpen
  \bibfield  {author} {\bibinfo {author} {\bibfnamefont {C.}~\bibnamefont
  {Amole}} \emph {et~al.} (\bibinfo {collaboration} {PICO}),\ }\href
  {https://doi.org/10.1103/PhysRevLett.118.251301} {\bibfield  {journal}
  {\bibinfo  {journal} {Phys. Rev. Lett.}\ }\textbf {\bibinfo {volume} {118}},\
  \bibinfo {pages} {251301} (\bibinfo {year} {2017})}\BibitemShut {NoStop}%
\bibitem [{\citenamefont {Meng}\ \emph {et~al.}(2021)\citenamefont {Meng} \emph
  {et~al.}}]{meng2021dark}%
  \BibitemOpen
  \bibfield  {author} {\bibinfo {author} {\bibfnamefont {Y.}~\bibnamefont
  {Meng}} \emph {et~al.} (\bibinfo {collaboration} {PandaX}),\ }\href@noop {}
  {\bibfield  {journal} {\bibinfo  {journal} {Phys. Rev. Lett.}\ }\textbf
  {\bibinfo {volume} {127}},\ \bibinfo {pages} {261802} (\bibinfo {year}
  {2021})}\BibitemShut {NoStop}%
\bibitem [{\citenamefont {Kolb}\ and\ \citenamefont
  {Turner}(1990)}]{Kolb:1990vq}%
  \BibitemOpen
  \bibfield  {author} {\bibinfo {author} {\bibfnamefont {E.~W.}\ \bibnamefont
  {Kolb}}\ and\ \bibinfo {author} {\bibfnamefont {M.~S.}\ \bibnamefont
  {Turner}},\ }\href {https://doi.org/10.1201/9780429492860} {\emph {\bibinfo
  {title} {{The Early Universe}}}},\ Vol.~\bibinfo {volume} {69}\ (\bibinfo
  {year} {1990})\BibitemShut {NoStop}%
\bibitem [{\citenamefont {Hall}\ \emph {et~al.}(2010)\citenamefont {Hall},
  \citenamefont {Jedamzik}, \citenamefont {March-Russell},\ and\ \citenamefont
  {West}}]{Hall:2009bx}%
  \BibitemOpen
  \bibfield  {author} {\bibinfo {author} {\bibfnamefont {L.~J.}\ \bibnamefont
  {Hall}}, \bibinfo {author} {\bibfnamefont {K.}~\bibnamefont {Jedamzik}},
  \bibinfo {author} {\bibfnamefont {J.}~\bibnamefont {March-Russell}},\ and\
  \bibinfo {author} {\bibfnamefont {S.~M.}\ \bibnamefont {West}},\ }\href
  {https://doi.org/10.1007/JHEP03(2010)080} {\bibfield  {journal} {\bibinfo
  {journal} {JHEP}\ }\textbf {\bibinfo {volume} {03}},\ \bibinfo {pages}
  {080}},\ \Eprint {https://arxiv.org/abs/0911.1120} {arXiv:0911.1120 [hep-ph]}
  \BibitemShut {NoStop}%
\bibitem [{\citenamefont {Cheng}\ \emph {et~al.}(2021)\citenamefont {Cheng}
  \emph {et~al.}}]{pandaxiiewimp}%
  \BibitemOpen
  \bibfield  {author} {\bibinfo {author} {\bibfnamefont {C.}~\bibnamefont
  {Cheng}} \emph {et~al.} (\bibinfo {collaboration} {PandaX}),\ }\href
  {https://doi.org/10.1103/PhysRevLett.126.211803} {\bibfield  {journal}
  {\bibinfo  {journal} {Phys. Rev. Lett.}\ }\textbf {\bibinfo {volume} {126}},\
  \bibinfo {pages} {211803} (\bibinfo {year} {2021})}\BibitemShut {NoStop}%
\bibitem [{\citenamefont {Essig}\ \emph
  {et~al.}(2012{\natexlab{a}})\citenamefont {Essig}, \citenamefont {Mardon},\
  and\ \citenamefont {Volansky}}]{PhysRevD.85.076007DMmodel}%
  \BibitemOpen
  \bibfield  {author} {\bibinfo {author} {\bibfnamefont {R.}~\bibnamefont
  {Essig}}, \bibinfo {author} {\bibfnamefont {J.}~\bibnamefont {Mardon}},\ and\
  \bibinfo {author} {\bibfnamefont {T.}~\bibnamefont {Volansky}},\ }\href
  {https://doi.org/10.1103/PhysRevD.85.076007} {\bibfield  {journal} {\bibinfo
  {journal} {Phys. Rev. D}\ }\textbf {\bibinfo {volume} {85}},\ \bibinfo
  {pages} {076007} (\bibinfo {year} {2012}{\natexlab{a}})}\BibitemShut
  {NoStop}%
\bibitem [{\citenamefont {Essig}\ \emph
  {et~al.}(2012{\natexlab{b}})\citenamefont {Essig}, \citenamefont
  {Manalaysay}, \citenamefont {Mardon}, \citenamefont {Sorensen},\ and\
  \citenamefont {Volansky}}]{PhysRevLett.109.021301}%
  \BibitemOpen
  \bibfield  {author} {\bibinfo {author} {\bibfnamefont {R.}~\bibnamefont
  {Essig}}, \bibinfo {author} {\bibfnamefont {A.}~\bibnamefont {Manalaysay}},
  \bibinfo {author} {\bibfnamefont {J.}~\bibnamefont {Mardon}}, \bibinfo
  {author} {\bibfnamefont {P.}~\bibnamefont {Sorensen}},\ and\ \bibinfo
  {author} {\bibfnamefont {T.}~\bibnamefont {Volansky}},\ }\href
  {https://doi.org/10.1103/PhysRevLett.109.021301} {\bibfield  {journal}
  {\bibinfo  {journal} {Phys. Rev. Lett.}\ }\textbf {\bibinfo {volume} {109}},\
  \bibinfo {pages} {021301} (\bibinfo {year} {2012}{\natexlab{b}})}\BibitemShut
  {NoStop}%
\bibitem [{\citenamefont {Aguilar-Arevalo}\ \emph {et~al.}(2019)\citenamefont
  {Aguilar-Arevalo} \emph {et~al.}}]{PhysRevLett.123.181802}%
  \BibitemOpen
  \bibfield  {author} {\bibinfo {author} {\bibfnamefont {A.}~\bibnamefont
  {Aguilar-Arevalo}} \emph {et~al.} (\bibinfo {collaboration} {DAMIC}),\ }\href
  {https://doi.org/10.1103/PhysRevLett.123.181802} {\bibfield  {journal}
  {\bibinfo  {journal} {Phys. Rev. Lett.}\ }\textbf {\bibinfo {volume} {123}},\
  \bibinfo {pages} {181802} (\bibinfo {year} {2019})}\BibitemShut {NoStop}%
\bibitem [{\citenamefont {Emken}\ \emph {et~al.}(2019)\citenamefont {Emken},
  \citenamefont {Essig}, \citenamefont {Kouvaris},\ and\ \citenamefont
  {Sholapurkar}}]{Emken_2019}%
  \BibitemOpen
  \bibfield  {author} {\bibinfo {author} {\bibfnamefont {T.}~\bibnamefont
  {Emken}}, \bibinfo {author} {\bibfnamefont {R.}~\bibnamefont {Essig}},
  \bibinfo {author} {\bibfnamefont {C.}~\bibnamefont {Kouvaris}},\ and\
  \bibinfo {author} {\bibfnamefont {M.}~\bibnamefont {Sholapurkar}},\ }\href
  {https://doi.org/10.1088/1475-7516/2019/09/070} {\bibfield  {journal}
  {\bibinfo  {journal} {Journal of Cosmology and Astroparticle Physics}\
  }\textbf {\bibinfo {volume} {2019}}\bibinfo  {number} { (09)},\ \bibinfo
  {pages} {070}}\BibitemShut {NoStop}%
\bibitem [{\citenamefont {Essig}\ \emph {et~al.}(2016)\citenamefont {Essig},
  \citenamefont {Fernandez-Serra}, \citenamefont {Mardon}, \citenamefont
  {Soto}, \citenamefont {Volansky},\ and\ \citenamefont {Yu}}]{Essig:2015cda}%
  \BibitemOpen
\bibfield  {number} {  }\bibfield  {author} {\bibinfo {author} {\bibfnamefont
  {R.}~\bibnamefont {Essig}}, \bibinfo {author} {\bibfnamefont
  {M.}~\bibnamefont {Fernandez-Serra}}, \bibinfo {author} {\bibfnamefont
  {J.}~\bibnamefont {Mardon}}, \bibinfo {author} {\bibfnamefont
  {A.}~\bibnamefont {Soto}}, \bibinfo {author} {\bibfnamefont {T.}~\bibnamefont
  {Volansky}},\ and\ \bibinfo {author} {\bibfnamefont {T.-T.}\ \bibnamefont
  {Yu}},\ }\href {https://doi.org/10.1007/JHEP05(2016)046} {\bibfield
  {journal} {\bibinfo  {journal} {JHEP}\ }\textbf {\bibinfo {volume} {05}},\
  \bibinfo {pages} {046}},\ \Eprint {https://arxiv.org/abs/1509.01598}
  {arXiv:1509.01598 [hep-ph]} \BibitemShut {NoStop}%
\bibitem [{\citenamefont {Essig}\ \emph
  {et~al.}(2017{\natexlab{a}})\citenamefont {Essig}, \citenamefont {Volansky},\
  and\ \citenamefont {Yu}}]{Essig:2017kqs}%
  \BibitemOpen
  \bibfield  {author} {\bibinfo {author} {\bibfnamefont {R.}~\bibnamefont
  {Essig}}, \bibinfo {author} {\bibfnamefont {T.}~\bibnamefont {Volansky}},\
  and\ \bibinfo {author} {\bibfnamefont {T.-T.}\ \bibnamefont {Yu}},\ }\href
  {https://doi.org/10.1103/PhysRevD.96.043017} {\bibfield  {journal} {\bibinfo
  {journal} {Phys. Rev. D}\ }\textbf {\bibinfo {volume} {96}},\ \bibinfo
  {pages} {043017} (\bibinfo {year} {2017}{\natexlab{a}})},\ \Eprint
  {https://arxiv.org/abs/1703.00910} {arXiv:1703.00910 [hep-ph]} \BibitemShut
  {NoStop}%
\bibitem [{\citenamefont {Zhang}\ \emph {et~al.}(2019)\citenamefont {Zhang}
  \emph {et~al.}}]{hongguang}%
  \BibitemOpen
  \bibfield  {author} {\bibinfo {author} {\bibfnamefont {H.}~\bibnamefont
  {Zhang}} \emph {et~al.} (\bibinfo {collaboration} {PandaX}),\ }\href
  {https://doi.org/10.1007/s11433-018-9259-0} {\bibfield  {journal} {\bibinfo
  {journal} {Sci. China Phys. Mech. Astron.}\ }\textbf {\bibinfo {volume}
  {62}},\ \bibinfo {pages} {31011} (\bibinfo {year} {2019})}\BibitemShut
  {NoStop}%
\bibitem [{\citenamefont {Qian}\ \emph {et~al.}(2022)\citenamefont {Qian} \emph
  {et~al.}}]{Qian_2022}%
  \BibitemOpen
  \bibfield  {author} {\bibinfo {author} {\bibfnamefont {Z.}~\bibnamefont
  {Qian}} \emph {et~al.} (\bibinfo {collaboration} {PandaX}),\ }\bibfield
  {journal} {\bibinfo  {journal} {Journal of High Energy Physics}\ }\textbf
  {\bibinfo {volume} {147}},\ \href {https://doi.org/10.1007/JHEP06(2022)147}
  {10.1007/JHEP06(2022)147} (\bibinfo {year} {2022})\BibitemShut {NoStop}%
\bibitem [{\citenamefont {He}\ \emph {et~al.}(2021)\citenamefont {He},
  \citenamefont {Liu}, \citenamefont {Ren}, \citenamefont {Shang},
  \citenamefont {Wei}, \citenamefont {Wang}, \citenamefont {Yang},
  \citenamefont {Yang}, \citenamefont {Yang}, \citenamefont {Zhang},\ and\
  \citenamefont {Zheng}}]{He_2021}%
  \BibitemOpen
  \bibfield  {author} {\bibinfo {author} {\bibfnamefont {C.}~\bibnamefont
  {He}}, \bibinfo {author} {\bibfnamefont {J.}~\bibnamefont {Liu}}, \bibinfo
  {author} {\bibfnamefont {X.}~\bibnamefont {Ren}}, \bibinfo {author}
  {\bibfnamefont {X.}~\bibnamefont {Shang}}, \bibinfo {author} {\bibfnamefont
  {X.}~\bibnamefont {Wei}}, \bibinfo {author} {\bibfnamefont {M.}~\bibnamefont
  {Wang}}, \bibinfo {author} {\bibfnamefont {J.}~\bibnamefont {Yang}}, \bibinfo
  {author} {\bibfnamefont {J.}~\bibnamefont {Yang}}, \bibinfo {author}
  {\bibfnamefont {Y.}~\bibnamefont {Yang}}, \bibinfo {author} {\bibfnamefont
  {G.}~\bibnamefont {Zhang}},\ and\ \bibinfo {author} {\bibfnamefont
  {Q.}~\bibnamefont {Zheng}},\ }\href
  {https://doi.org/10.1088/1748-0221/16/12/t12015} {\bibfield  {journal}
  {\bibinfo  {journal} {Journal of Instrumentation}\ }\textbf {\bibinfo
  {volume} {16}}\bibinfo  {number} { (12)},\ \bibinfo {pages}
  {T12015}}\BibitemShut {NoStop}%
\bibitem [{\citenamefont {Chen}\ \emph {et~al.}(2021)\citenamefont {Chen},
  \citenamefont {Cheng}, \citenamefont {Fu}, \citenamefont {Giuliani},
  \citenamefont {Liu}, \citenamefont {Lu}, \citenamefont {Ji}, \citenamefont
  {Qian}, \citenamefont {Qiao}, \citenamefont {Wang}, \citenamefont {Xia},
  \citenamefont {Xie}, \citenamefont {Yao},\ and\ \citenamefont
  {Zhang}}]{Chen_2021}%
  \BibitemOpen
\bibfield  {number} {  }\bibfield  {author} {\bibinfo {author} {\bibfnamefont
  {X.}~\bibnamefont {Chen}}, \bibinfo {author} {\bibfnamefont {C.}~\bibnamefont
  {Cheng}}, \bibinfo {author} {\bibfnamefont {M.}~\bibnamefont {Fu}}, \bibinfo
  {author} {\bibfnamefont {F.}~\bibnamefont {Giuliani}}, \bibinfo {author}
  {\bibfnamefont {J.}~\bibnamefont {Liu}}, \bibinfo {author} {\bibfnamefont
  {X.}~\bibnamefont {Lu}}, \bibinfo {author} {\bibfnamefont {X.}~\bibnamefont
  {Ji}}, \bibinfo {author} {\bibfnamefont {Z.}~\bibnamefont {Qian}}, \bibinfo
  {author} {\bibfnamefont {H.}~\bibnamefont {Qiao}}, \bibinfo {author}
  {\bibfnamefont {Q.}~\bibnamefont {Wang}}, \bibinfo {author} {\bibfnamefont
  {J.}~\bibnamefont {Xia}}, \bibinfo {author} {\bibfnamefont {P.}~\bibnamefont
  {Xie}}, \bibinfo {author} {\bibfnamefont {Y.}~\bibnamefont {Yao}},\ and\
  \bibinfo {author} {\bibfnamefont {H.}~\bibnamefont {Zhang}},\ }\href
  {https://doi.org/10.1088/1748-0221/16/09/t09004} {\bibfield  {journal}
  {\bibinfo  {journal} {Journal of Instrumentation}\ }\textbf {\bibinfo
  {volume} {16}}\bibinfo  {number} { (09)},\ \bibinfo {pages}
  {T09004}}\BibitemShut {NoStop}%
\bibitem [{\citenamefont {Zhao}\ \emph {et~al.}(2021)\citenamefont {Zhao},
  \citenamefont {Cui}, \citenamefont {Ma}, \citenamefont {Fan}, \citenamefont
  {Giboni}, \citenamefont {Zhang}, \citenamefont {Liu},\ and\ \citenamefont
  {Ji}}]{Zhao_2021}%
  \BibitemOpen
\bibfield  {number} {  }\bibfield  {author} {\bibinfo {author} {\bibfnamefont
  {L.}~\bibnamefont {Zhao}}, \bibinfo {author} {\bibfnamefont {X.}~\bibnamefont
  {Cui}}, \bibinfo {author} {\bibfnamefont {W.}~\bibnamefont {Ma}}, \bibinfo
  {author} {\bibfnamefont {Y.}~\bibnamefont {Fan}}, \bibinfo {author}
  {\bibfnamefont {K.}~\bibnamefont {Giboni}}, \bibinfo {author} {\bibfnamefont
  {T.}~\bibnamefont {Zhang}}, \bibinfo {author} {\bibfnamefont
  {J.}~\bibnamefont {Liu}},\ and\ \bibinfo {author} {\bibfnamefont
  {X.}~\bibnamefont {Ji}},\ }\href
  {https://doi.org/10.1088/1748-0221/16/06/t06007} {\bibfield  {journal}
  {\bibinfo  {journal} {Journal of Instrumentation}\ }\textbf {\bibinfo
  {volume} {16}}\bibinfo  {number} { (06)},\ \bibinfo {pages}
  {T06007}}\BibitemShut {NoStop}%
\bibitem [{\citenamefont {Wang}\ \emph {et~al.}(2021)\citenamefont {Wang},
  \citenamefont {Li}, \citenamefont {Ju}, \citenamefont {Lei}, \citenamefont
  {Liu}, \citenamefont {Ji}, \citenamefont {Tang},\ and\ \citenamefont
  {Gou}}]{XiuliIndium2021}%
  \BibitemOpen
\bibfield  {number} {  }\bibfield  {author} {\bibinfo {author} {\bibfnamefont
  {X.}~\bibnamefont {Wang}}, \bibinfo {author} {\bibfnamefont {S.}~\bibnamefont
  {Li}}, \bibinfo {author} {\bibfnamefont {Y.}~\bibnamefont {Ju}}, \bibinfo
  {author} {\bibfnamefont {Z.}~\bibnamefont {Lei}}, \bibinfo {author}
  {\bibfnamefont {J.}~\bibnamefont {Liu}}, \bibinfo {author} {\bibfnamefont
  {X.}~\bibnamefont {Ji}}, \bibinfo {author} {\bibfnamefont {X.}~\bibnamefont
  {Tang}},\ and\ \bibinfo {author} {\bibfnamefont {Y.}~\bibnamefont {Gou}},\
  }\href {https://doi.org/10.1063/5.0051279} {\bibfield  {journal} {\bibinfo
  {journal} {Review of Scientific Instruments}\ }\textbf {\bibinfo {volume}
  {92}},\ \bibinfo {pages} {093905} (\bibinfo {year} {2021})}\BibitemShut
  {NoStop}%
\bibitem [{\citenamefont {Ma}\ \emph {et~al.}()\citenamefont {Ma} \emph
  {et~al.}}]{b8_paper}%
  \BibitemOpen
  \bibfield  {author} {\bibinfo {author} {\bibfnamefont {W.}~\bibnamefont {Ma}}
  \emph {et~al.} (\bibinfo {collaboration} {PandaX}),\ }\href@noop {} {\
  }\Eprint {https://arxiv.org/abs/2207.04883} {arXiv:2207.04883} \BibitemShut
  {NoStop}%
\bibitem [{Note1()}]{Note1}%
  \BibitemOpen
  \bibinfo {note} {Such background can also emerge from the gate electrode, but
  is significantly suppressed by our data selection based on the $S2$
  width}\BibitemShut {NoStop}%
\bibitem [{Note2()}]{Note2}%
  \BibitemOpen
  \bibinfo {note} {Under cut1, $S2$s in set 2 are stretched due to a slightly
  different $S2$ gain of this dataset.}\BibitemShut {Stop}%
\bibitem [{\citenamefont {Aprile}\ \emph {et~al.}(2019)\citenamefont {Aprile}
  \emph {et~al.}}]{aprile2019light}%
  \BibitemOpen
  \bibfield  {author} {\bibinfo {author} {\bibfnamefont {E.}~\bibnamefont
  {Aprile}} \emph {et~al.} (\bibinfo {collaboration} {XENON}),\ }\href@noop {}
  {\bibfield  {journal} {\bibinfo  {journal} {Phys. Rev. Lett.}\ }\textbf
  {\bibinfo {volume} {123}},\ \bibinfo {pages} {251801} (\bibinfo {year}
  {2019})}\BibitemShut {NoStop}%
\bibitem [{\citenamefont {Agnes}\ \emph
  {et~al.}(2022{\natexlab{a}})\citenamefont {Agnes} \emph
  {et~al.}}]{agnes2022er}%
  \BibitemOpen
  \bibfield  {author} {\bibinfo {author} {\bibfnamefont {P.}~\bibnamefont
  {Agnes}} \emph {et~al.} (\bibinfo {collaboration} {DarkSide-50}),\
  }\href@noop {} {\  (\bibinfo {year} {2022}{\natexlab{a}})},\ \Eprint
  {https://arxiv.org/abs/2207.11968} {arXiv:2207.11968} \BibitemShut {NoStop}%
\bibitem [{\citenamefont {Essig}\ \emph
  {et~al.}(2017{\natexlab{b}})\citenamefont {Essig}, \citenamefont {Volansky},\
  and\ \citenamefont {Yu}}]{essig2017new}%
  \BibitemOpen
  \bibfield  {author} {\bibinfo {author} {\bibfnamefont {R.}~\bibnamefont
  {Essig}}, \bibinfo {author} {\bibfnamefont {T.}~\bibnamefont {Volansky}},\
  and\ \bibinfo {author} {\bibfnamefont {T.-T.}\ \bibnamefont {Yu}},\
  }\href@noop {} {\bibfield  {journal} {\bibinfo  {journal} {Phys. Rev. D}\
  }\textbf {\bibinfo {volume} {96}},\ \bibinfo {pages} {043017} (\bibinfo
  {year} {2017}{\natexlab{b}})}\BibitemShut {NoStop}%
\bibitem [{\citenamefont {Abramoff}\ \emph {et~al.}(2019)\citenamefont
  {Abramoff} \emph {et~al.}}]{Abramoff_2019}%
  \BibitemOpen
  \bibfield  {author} {\bibinfo {author} {\bibfnamefont {O.}~\bibnamefont
  {Abramoff}} \emph {et~al.} (\bibinfo {collaboration} {SENSEI}),\ }\href
  {https://doi.org/10.1103/PhysRevLett.122.161801} {\bibfield  {journal}
  {\bibinfo  {journal} {Phys. Rev. Lett.}\ }\textbf {\bibinfo {volume} {122}},\
  \bibinfo {pages} {161801} (\bibinfo {year} {2019})}\BibitemShut {NoStop}%
\bibitem [{\citenamefont {Aprile}\ \emph {et~al.}(2021)\citenamefont {Aprile}
  \emph {et~al.}}]{aprile2021search}%
  \BibitemOpen
  \bibfield  {author} {\bibinfo {author} {\bibfnamefont {E.}~\bibnamefont
  {Aprile}} \emph {et~al.} (\bibinfo {collaboration} {XENON}),\ }\href@noop {}
  {\bibfield  {journal} {\bibinfo  {journal} {Phys. Rev. Lett.}\ }\textbf
  {\bibinfo {volume} {126}},\ \bibinfo {pages} {091301} (\bibinfo {year}
  {2021})}\BibitemShut {NoStop}%
\bibitem [{\citenamefont {Akerib}\ \emph
  {et~al.}(2017{\natexlab{b}})\citenamefont {Akerib} \emph
  {et~al.}}]{akerib2017results}%
  \BibitemOpen
  \bibfield  {author} {\bibinfo {author} {\bibfnamefont {D.}~\bibnamefont
  {Akerib}} \emph {et~al.} (\bibinfo {collaboration} {LUX}),\ }\href@noop {}
  {\bibfield  {journal} {\bibinfo  {journal} {Phys. Rev. Lett.}\ }\textbf
  {\bibinfo {volume} {118}},\ \bibinfo {pages} {021303} (\bibinfo {year}
  {2017}{\natexlab{b}})}\BibitemShut {NoStop}%
\bibitem [{\citenamefont {Agnes}\ \emph
  {et~al.}(2022{\natexlab{b}})\citenamefont {Agnes} \emph
  {et~al.}}]{agnes2022_nr}%
  \BibitemOpen
  \bibfield  {author} {\bibinfo {author} {\bibfnamefont {P.}~\bibnamefont
  {Agnes}} \emph {et~al.} (\bibinfo {collaboration} {DarkSide-50}),\
  }\href@noop {} {\  (\bibinfo {year} {2022}{\natexlab{b}})},\ \Eprint
  {https://arxiv.org/abs/2207.11966} {arXiv:2207.11966} \BibitemShut {NoStop}%
\bibitem [{\citenamefont {Szydagis}\ \emph {et~al.}(2018)\citenamefont
  {Szydagis}, \citenamefont {Balajthy}, \citenamefont {Brodsky}, \citenamefont
  {Cutter}, \citenamefont {Huang}, \citenamefont {Kozlova}, \citenamefont
  {Lenardo}, \citenamefont {Manalaysay}, \citenamefont {McKinsey},
  \citenamefont {Mooney} \emph {et~al.}}]{szydagis2018noble}%
  \BibitemOpen
  \bibfield  {author} {\bibinfo {author} {\bibfnamefont {M.}~\bibnamefont
  {Szydagis}}, \bibinfo {author} {\bibfnamefont {J.}~\bibnamefont {Balajthy}},
  \bibinfo {author} {\bibfnamefont {J.}~\bibnamefont {Brodsky}}, \bibinfo
  {author} {\bibfnamefont {J.}~\bibnamefont {Cutter}}, \bibinfo {author}
  {\bibfnamefont {J.}~\bibnamefont {Huang}}, \bibinfo {author} {\bibfnamefont
  {E.}~\bibnamefont {Kozlova}}, \bibinfo {author} {\bibfnamefont
  {B.}~\bibnamefont {Lenardo}}, \bibinfo {author} {\bibfnamefont
  {A.}~\bibnamefont {Manalaysay}}, \bibinfo {author} {\bibfnamefont
  {D.}~\bibnamefont {McKinsey}}, \bibinfo {author} {\bibfnamefont
  {M.}~\bibnamefont {Mooney}}, \emph {et~al.},\ }\href@noop {} {\bibfield
  {journal} {\bibinfo  {journal} {Zenodo: Geneve, Switzerland}\ } (\bibinfo
  {year} {2018})}\BibitemShut {NoStop}%
\bibitem [{\citenamefont {Szydagis}(2022)}]{szydagis2022noble}%
  \BibitemOpen
  \bibfield  {author} {\bibinfo {author} {\bibfnamefont {M.}~\bibnamefont
  {Szydagis}},\ }\href@noop {} {\bibfield  {journal} {\bibinfo  {journal}
  {Bulletin of the American Physical Society}\ } (\bibinfo {year}
  {2022})}\BibitemShut {NoStop}%
\bibitem [{\citenamefont {Akerib}\ \emph
  {et~al.}(2017{\natexlab{c}})\citenamefont {Akerib} \emph
  {et~al.}}]{LUX:2017ojt}%
  \BibitemOpen
  \bibfield  {author} {\bibinfo {author} {\bibfnamefont {D.~S.}\ \bibnamefont
  {Akerib}} \emph {et~al.} (\bibinfo {collaboration} {LUX}),\ }\href
  {https://doi.org/10.1103/PhysRevD.96.112011} {\bibfield  {journal} {\bibinfo
  {journal} {Phys. Rev. D}\ }\textbf {\bibinfo {volume} {96}},\ \bibinfo
  {pages} {112011} (\bibinfo {year} {2017}{\natexlab{c}})},\ \Eprint
  {https://arxiv.org/abs/1709.00800} {arXiv:1709.00800 [physics.ins-det]}
  \BibitemShut {NoStop}%
\bibitem [{\citenamefont {Baxter}\ \emph {et~al.}(2021)\citenamefont {Baxter},
  \citenamefont {Bloch}, \citenamefont {Bodnia}, \citenamefont {Chen},
  \citenamefont {Conrad}, \citenamefont {Di~Gangi}, \citenamefont {Dobson},
  \citenamefont {Durnford}, \citenamefont {Haselschwardt}, \citenamefont
  {Kaboth} \emph {et~al.}}]{baxter2021recommended}%
  \BibitemOpen
  \bibfield  {author} {\bibinfo {author} {\bibfnamefont {D.}~\bibnamefont
  {Baxter}}, \bibinfo {author} {\bibfnamefont {I.}~\bibnamefont {Bloch}},
  \bibinfo {author} {\bibfnamefont {E.}~\bibnamefont {Bodnia}}, \bibinfo
  {author} {\bibfnamefont {X.}~\bibnamefont {Chen}}, \bibinfo {author}
  {\bibfnamefont {J.}~\bibnamefont {Conrad}}, \bibinfo {author} {\bibfnamefont
  {P.}~\bibnamefont {Di~Gangi}}, \bibinfo {author} {\bibfnamefont
  {J.}~\bibnamefont {Dobson}}, \bibinfo {author} {\bibfnamefont
  {D.}~\bibnamefont {Durnford}}, \bibinfo {author} {\bibfnamefont
  {S.}~\bibnamefont {Haselschwardt}}, \bibinfo {author} {\bibfnamefont
  {A.}~\bibnamefont {Kaboth}}, \emph {et~al.},\ }\href@noop {} {\bibfield
  {journal} {\bibinfo  {journal} {The European Physical Journal C}\ }\textbf
  {\bibinfo {volume} {81}},\ \bibinfo {pages} {1} (\bibinfo {year}
  {2021})}\BibitemShut {NoStop}%
\end{thebibliography}%
